\documentclass[epj]{svjour}
\usepackage{amsmath,graphicx,slashed,indentfirst,fixltx2e}

\newcommand{\ud}{\mathrm{d}}
\newcommand{\abs[1]}{\left| #1 \right|}
\newcommand{\ket}[1]{\left| #1 \right>}
\newcommand{\bra}[1]{\left< #1 \right|}
\begin{document}
\title{Off-shell photon distribution amplitudes in the low-energy effective theory of QCD}
\author{Xin Mo \and Jueping Liu\inst{}
\thanks{Author to whom all correspondences should be addressed.
E-mail address: jpliu@whu.edu.cn}%
}                     
%
%
\institute{Department of Physics,School of Physics Science and Technology,Wuhan University,430072,Wuhan,China}
\date{Received: date / Revised version: date}
%
\abstract{
Based on the principle of the Lorentz covariance the transition matrix elements from an off-shell photon state to the vacuum are decomposed into the light-cone photon DAs, in which only two transversal DAs survive in the on-shell limit. The eight off-shell light-cone photon distribution amplitudes (DAs) corresponding to chiral-odd and chiral-even up to twist-four and the corresponding coupling constants are studied systematically in the instanton vacuum model of quantum chromodynamics (QCD).
The various individual photon DA multiplied by its corresponding coupling constant is expressed in terms of the correlation functions, which are connected with the spectral densities of an effective quark propagator, and then evaluated in the low-energy effective theory derived from the instanton vacuum model of QCD.
The explicit analytical expressions and the numerical results for the photon DAs and their coupling constants are given.
\PACS{
      {PACS-key}{11.15.Tk,12.38.Lg,11.55.Hx,14.70.Bh}   
     } 
} 
\maketitle
\section{Introduction}
Motivated originally by Brodsky and Lepage\cite{QCD}, hadronic distribution amplitudes(DAs) are introduced as the nonperturbative parameters to deal with hadronic scattering process\cite{Lepage79PLB,Lepage79PRL1,Lepage79PRL2,Lepage79PRD,Lepage80,Cher84}. Similar to the hadronic ones, the photon DAs are also introduced into QCD light-cone sum rules\cite{Braun94}, serve as the reliable non-perturbative parameters in various processes involving photons, such as radiative decay of hyperons\cite{Bali89}, the scattering of real and virtual photons $\gamma\gamma^*\to\pi^0$\cite{Rady96NPB,Rady96PLB1}, and the deeply virtual Compton scattering\cite{Rady96PLB2,Rady96PLB3}.

For hadronic DAs, such as those associated with pion\cite{Pra01,Pra02}, rho\cite{Ball98} and  kaon\cite{Nam06prd1,Nam06prd2,Nam06prd3}, huge amount of works have been done in literatures. In comparison, there are only a few works for calculation of the photon DAs yet, especially for the ones of higher twist, based on a consistent formalism.

To the best of our knowledge, the most detailed treatment on photon DAs of the asymptotic form has been carried out in \cite{npb2003photon}. It provides a systemic classification of photon DAs corresponding to different twists, number of intermediate particles and the chiralities in the framework of the background field formalism. However, it deals simply with the real photon case. We understand these results as being valid in the high-energy region, because the conformal symmetry is used as a base for expansion. Then, some extended work have been done by means of both an asymptotic treatment and the effective low-energy theory\cite{prd2006,prd2010}. A very impressive common characteristic of the results is that a scalar type of the virtual photon DA is missing. The reason may be traced back to the fact that the authors have adopted two mixing schemes: the background field formalism\cite{npb2003photon} which may be understood to be valid in high-energy region, and the instanton vacuum model which is basically appropriate at the low-energy region\cite{prd2006,prd2010}.

To exploit such point, we completely confine ourself in the low-energy effective theory of the local type, derived from the instanton vacuum model of QCD. Along this line, the leading twist light-cone real photon DA corresponding to the tensor current has been calculated\cite{prd1999pi}. Within this theoretical framework, the QCD vacuum is described as a dilute medium of instantons, the interaction of quarks with the fermion zero modes of the individual instanton in this medium provides a mechanism for the chiral symmetry breaking\cite{Didko96}, which is the most important non-perturbative phenomenon in the hadron world. The calculated masses and coupling constants of hadrons, such as $\pi$, $\sigma$, $\rho$, $A_1$, $N$, $\Delta$, and etc., agree with data, and the corresponding correlation functions are also in accordance with phenomenology\cite{Shur93} and lattice simulations\cite{Chu93,Tsch96}. The picture of the instanton vacuum also leads to the formation of the gluon condensates\cite{Shi79,Liu93} and the so-called topological susceptibility needed to cure the $U(1)$ paradox\cite{tHoo76prl,tHoo76prd}. At the large-$N_c$ limit, the instanton vacuum leads to a very reasonable low-energy effective theory of quarks with a momentum-dependent dynamical mass $M(k)$, which drops to zero at Euclidean momentum of the order of the inverse of the average instanton size, $\bar{\rho}^{-1}\approx 600$MeV\cite{quark}. The whole approach of this low-energy effective theory is based on the smallness of $(M(k)\bar{\rho})^2$, and is restricted to momenta $k\ll 1/\bar{\rho}$, at such low-energy scale there are no dynamical gluons.

Following the same line as in \cite{prd1999pi}, we have calculated the off-shell photon DAs at leading twist corresponding with both Dirac structures, $\sigma_{\mu\nu}$ and $\gamma_{\mu}$\cite{yuran06prd,yuran06EPJA,zhukai06}, and the twist-two parts of the other two photon DAs (each of which contains both twist-two and twist-three parts, but we call them as twist-three DAs in comparison with other references) of the tensor current are estimated.

In the present paper, we extend our previous calculation to higher twists, and give a complete treatment in the same way for all light-cone photon DAs up to twist-four appearing in the formalism described before in \cite{yuran06prd}, which is based on the effective low-energy instanton vacuum model of QCD and a systematical Lorentz decomposition of the transition matrix element from an off-shell photon to vacuum through a nonlocal quark-antiquark bilinear current with a gauge link. We show that in our formalism there are altogether eight photon DAs up to twist-four, in which only two DAs survive in the on-shell limit. In particular, there appears a scalar virtual photon DA corresponding to the scalar nonlocal bilinear quark current, as explicitly demonstrated in calculation. However, such DA is absent in the asymptotic form because, as the authors claimed, the vacuum expectation value of the scalar operator does not contain any contribution linear in the background electromagnetic field.

Besides the above mentioned difference, there is another difference between our treatment and the one used in Refs. \cite{prd2006,prd2010}.
It is noted that when working with the effective theory (the covariant version of \eqref{eq:31} below), the dynamical quark mass is assumed to be small compared with the ultraviolet cutoff. In order to compute the photon DAs  we need to couple an electromagnetic field to the quark fields of the effective Lagrangian. The interaction between the electromagnetic field and quarks should be dominated by the pointlike one, while the non-pointlike interaction, which arises from the momentum-dependent mass term, is suppressed in $(M\bar{\rho})^2$ relative to the pointlike one. Therefore, keeping in mind within leading order in $M\bar{\rho}$, we may choose to work with the pointlike electromagnetic current (a further discussion of this point can be seen the last paragraph and the Appendix C of this paper).

Our paper is organized as follows: After given our ideas and motivation in the introduction,
in Sec. \uppercase\expandafter{\romannumeral2} we review our previous formalism to define the off-shell light-cone photon DAs up to twist-four with a slight modification based just on the Lorentz covariance. Then, every photon DA multiplied by its coupling constant is individually expressed in terms of the corresponding correlation function in Sec. \uppercase\expandafter{\romannumeral3}. Afterwards, in Sec. \uppercase\expandafter{\romannumeral4} the spectral representations of the correlation functions are worked out by using a general spectral representation of the effective quark propagator, which is assumed to be of the pole form related to the instanton vacuum model of QCD for practical purpose. In Sec. \uppercase\expandafter{\romannumeral5}, the analytical expressions of all photon DAs are shown. The numerical simulation and the corresponding results are displayed in Sec. \uppercase\expandafter{\romannumeral6}. Finally, in Sec. \uppercase\expandafter{\romannumeral7} our conclusions and discussions are given. Some technical details are presented in the Appendixes.

\section{Definition of photon distribution amplitudes up to twist-four}
Let's review the formalism suggested in \cite{yuran06prd} with a slight modification, in which the photon DAs are defined, and classified into two groups with different chiralities. According to Lorentz covariance, the nonlocal quark-antiquark current with light-like separation ($z^2=0$) sandwiched between vacuum $\ket{0}$ and one-photon state $\ket{\gamma(P,\lambda)}$, characterized by its momentum $P_{\mu}$ and the polarization vector $e^{(\lambda)}_{\sigma}$, can be decomposed into different
Lorentz structures $L_i(p,n,e^{(\lambda)})$,
\begin{align}\label{eq:1}
\bra{0}&\bar{\psi}(z)\Gamma\left[z,-z\right]\psi(-z)\ket{\gamma(P,\lambda)}\nonumber\\
&=\sum_if_i(P^2)L_i(p,n,e^{(\lambda)})\int^1_0\ud u e^{i\xi p\cdot z}\phi^{(i)}_{\gamma}(u,P^2),
\end{align}
where $\Gamma$ is one of the Dirac-matrices $I$, $\gamma_{\mu}$, $\sigma_{\mu\nu}$, $\gamma_{\mu}\gamma_5$ and $\gamma_5$, $u$ the fraction of the momentum carried by the quark, $\xi =2u-1$, $\phi^{(i)}_{\gamma}(u,P^2)$ the photon DA with a virtuality $P^2$ being normalized as
\begin{eqnarray}\label{eq:2}
\int^1_0\ud u\phi^{(i)}_{\gamma}(u,P^2)=1.
\end{eqnarray}
and $f_i(P^2)$ its corresponding coupling constant. The notation $\left[x,y\right]$ is the path-ordered gauge link
\begin{eqnarray*}
\left[x,y\right]=&P\mathrm{exp}\left\{i\int^1_0\ud\tau(x-y)_{\mu}[g_sB^{\mu}(\tau x+(1-\tau)y)\right.\nonumber\\
&\left.+Q A^{\mu}(\tau x+(1-\tau)y)]\right\},
\end{eqnarray*}
with $B^{\mu}$ and $A^{\mu}$ being the gauge potentials for strong and electromagnetic interactions respectively, and $g_s$ and $Q$ the corresponding coupling constants.

To find the Lorentz structures $L_i(p,n,e^{(\lambda)})$ for $i=t,v$, corresponding to $\Gamma$ being $\sigma_{\mu\nu}$ (tensor case) and $\gamma_{\mu}$ (vector case) respectively, let us start with the local quark-antiquark bilinear case,  $z=0$. It is easy to see that there are only two physical
vectors, the polarization vector $e^{(\lambda)}_{\mu}$ and the momentum $P_{\nu}$, available, and each photon DA  should be linear in $e^{(\lambda)}_{\mu}$. Thus
\begin{eqnarray}
&\bra{0}\bar{\psi}(0)\sigma_{\mu\nu}\psi(0)\ket{\gamma(P,\lambda)}
=if^{(t)}_{\gamma}(P^2)T^{(\lambda)}_{\mu\nu}\label{eq:3}\\
&\bra{0}\bar{\psi}(0)\gamma_{\mu}\psi(0)\ket{\gamma(P,\lambda)}=M f^{(v)}_{\gamma}(P^2)e^{(\lambda)}_{\mu}\label{eq:4}
\end{eqnarray}
where $T^{(\lambda)}_{\mu\nu}=e^{(\lambda)}_{\mu}P_{\nu}-e^{(\lambda)}_{\nu}P_{\mu}$, and $M$ is a Lorentz invariant constant which is taken to be the effective non-vanishing quark mass at $P^2=0$ instead of $\mu\equiv \sqrt{\abs[P^2]}$ as before to avoid the influence of some inconvenient behavior of $\mu$ near $P^2=0$.

For implementing the light-cone expansion in a systematical way, let us introduce a light-like vector $p$ such that $P\rightarrow p$ as $P^2\rightarrow
0$; and for the sake of convenience, introduce further the dimensionless light-like vectors $n$ and $\hat{n}$ parallel to $z$ and $p$ respectively with
\begin{displaymath}
 n\cdot \hat{n}=2,\,\,\,\,\,z_{\mu}=n_{\mu}\tau
\end{displaymath}
Then, we go away from $z=0$ but keeping $z^2=0$, and decompose $e^{(\lambda)}_{\mu}$ and $P_{\mu}$ into
three independent vectors, $p_{\mu}$, $z_{\mu}$ and $e^{(\lambda)}_{\perp\mu}$ (the projection of $e^{(\lambda)}_{\mu}$
on the plane perpendicular to both $p_{\mu}$ and $z_{\mu}$), namely
\begin{align}
&e^{(\lambda)}_{\mu}
=e^{(\lambda)}_{\perp\mu}+p_{\mu}\frac{e^{(\lambda)}\cdot z}{p\cdot z}-z_{\mu}\frac{e^{(\lambda)}\cdot z}{2(p\cdot z)^2}P^2\label{eq:5}\\
&P_{\mu}=p_{\mu}+\frac{P^2}{2p\cdot z}z_{\mu}\label{eq:6}
\end{align}
which leads to
\begin{eqnarray}
T^{(\lambda)}_{\mu\nu}
=\sum_{i=1}^3T^{(\lambda,i)}_{\mu\nu} \label{eq:7}
\end{eqnarray}
with
\begin{align}
&T^{(\lambda,1)}_{\mu\nu}=
e_{\perp\mu}^{(\lambda)}p_{\nu}-e_{\perp\nu}^{(\lambda)}p_{\mu} \label{eq:8}\\
&T^{(\lambda,2)}_{\mu\nu}=(p_{\mu}n_{\nu}-p_{\nu}n_{\mu})
\frac{e^{(\lambda)}\cdot n}{(p\cdot n)^2} \label{eq:9}\\
&T^{(\lambda,3)}_{\mu\nu}=(e_{\perp\mu}^{(\lambda)}n_{\nu}
-e_{\perp\nu}^{(\lambda)}n_{\mu})\frac{P^2}{2p\cdot n} \label{eq:10}
\end{align}
We note here that because of the conservation of the local electromagnetic current, there is the orthogonality relation between the polarization vector and the momentum of a photon
\begin{displaymath}
e^{(\lambda)}\cdot P=0
\end{displaymath}
which can be used to transform $e^{(\lambda)}\cdot z$ into $e^{(\lambda)}\cdot p$, and vice versa, and to
obtain a formula for the photon polarization summation,
\begin{equation}
\sum_{\lambda}e^{*(\lambda)}_{\mu}e^{(\lambda)}_{\nu} = -g_{\mu\nu}+\frac{P_{\mu}P_{\nu}}{P^2}\label{eq:11}
\end{equation}

It is important to note that the three terms of the R.H.S. of \eqref{eq:8}-\eqref{eq:10} are the only independent antisymmetric tensors, which can be constructed from the three independent vectors $p_{\mu}$, $z_{\mu}$ and $e^{(\lambda)}_{\perp\mu}$.
As a consequence, we have the definition of the photon DAs for $z\not=0$
\begin{align}\label{eq:12}
\bra{0}&\bar{\psi}(z)\sigma_{\mu\nu}\left[z,-z\right]\psi(-z)\ket{\gamma(P,\lambda)}\nonumber\\
&=if^{(t)}_{\gamma\perp}(P^2)T^{(\lambda 1)}_{\mu\nu}\int^1_0\ud u e^{i\xi p\cdot z}\phi_{\gamma\perp}^{(t)}(u,P^2)\nonumber\\
&+if^{(t)}_{\gamma\parallel}(P^2)T^{(\lambda 2)}_{\mu\nu}\int^1_0\ud u e^{i\xi p\cdot z}h_{\gamma\parallel}^{(t)}(u,P^2)\nonumber\\
&+if^{(t)}_{\gamma3}(P^2)T^{(\lambda 3)}_{\mu\nu}\int^1_0\ud u e^{i\xi p\cdot z}h_{\gamma3}^{(t)}(u,P^2),\nonumber\\
\end{align}
for the tensor case, and
\begin{align}\label{eq:13}
&\bra{0}\bar{\psi}(z)\gamma_{\mu}\left[z,-z\right]\psi(-z)\ket{\gamma(P,\lambda)}\nonumber\\
&=f^{(v)}_{\gamma\parallel}(P^2)M p_{\mu}\frac{e^{(\lambda)}\cdot n}{p\cdot n}\int^1_0\ud u e^{i\xi p\cdot z}\phi_{\gamma\parallel}^{(v)}(u,P^2)\nonumber\\
&+f^{(v)}_{\gamma\perp}(P^2)M e_{\perp\mu}^{(\lambda)}\int^1_0\ud u e^{i\xi p\cdot z}g_{\gamma\perp}^{(v)}(u,P^2)\nonumber\\
&-f^{(v)}_{\gamma3}(P^2)M n_{\mu}\frac{e^{(\lambda)}\cdot n}{(p\cdot n)^2}P^2\int^1_0\ud u e^{i\xi p\cdot z}g_{\gamma3}^{(v)}(u,P^2)
\end{align}
for the vector case, according to the Lorentz decomposition \eqref{eq:1}.

In addition, it is obvious that there is no pseudoscalar which can be constructed from the three independent
vectors $p_{\mu}$, $z_{\mu}$ and $e^{(\lambda)}_{\perp\mu}$, and that only one scalar and one axial vector can be constructed from the above three independent vectors, and thus the corresponding photon DAs can be defined as follows
\begin{align}\label{eq:14}
\bra{0}&\bar{\psi}(z)\left[z,-z\right]\psi(-z)\ket{\gamma(P,\lambda)}\nonumber\\
&=if^{(s)}_{\gamma\parallel}(P^2)(e^{(\lambda)}\cdot z)P^2\int^1_0\ud u e^{i\xi p\cdot z}h^{(s)}_{\gamma\parallel}(u,P^2),
\end{align}
\begin{align}\label{eq:15}
\bra{0}&\bar{\psi}(z)\gamma_{\mu}\gamma_5\left[z,-z\right]\psi(-z)\ket{\gamma(P,\lambda)}\nonumber\\
&=M f^{(a)}_{\gamma\perp}(P^2)\epsilon_{\mu\nu\alpha\beta}e^{(\lambda)\nu}_{\perp}p^{\alpha}z^{\beta}\int^1_0\ud u e^{i\xi p\cdot z}g^{(a)}_{\gamma\perp}(u,P^2).
\end{align}
From equation \eqref{eq:14}, it is obvious that a scalar operator $\bar{\psi}(z)\left[z,-z\right]\psi(-z)$ has a projection onto a virtual photon state $\ket{\gamma(P,\lambda)}$ which is proportional to the longitudinal photon polarization $e^{(\lambda)}\cdot z$, and vanishing in the on-shell limit.

The above formalism for defining the light-cone photon DAs (or wave functions) is, in fact, parallel to the case of $\rho$ meson\cite{Ball98}. An obvious advantage of this formalism is that only two transversal DAs
$\phi^{(t)}_{\gamma\perp}(u,P^2)$ and $g^{(a)}_{\gamma\perp}(u,P^2)$ survive in the on-shell case, and the others decouple automatically from the corresponding quark-antiquark currents, provided all coupling constants are finite at $P^2=0$ as they should be (see below). This is just as we expected, since a real photon has only two transverse degrees of freedom.

\section{Photon DAs expressed in terms of correlation functions}
To evaluate the transition matrix elements between photon and vacuum for various currents,
we rewrite them as\cite{npb2003photon}
\begin{align}\label{eq:16}
\bra{0}&\bar{\psi}(z)\Gamma\left[z,-z\right]\psi(-z)\ket{\gamma(P,\lambda)}\nonumber\\
=&i\int\ud^4x e^{-iP\cdot x}e^{(\lambda)}_{\sigma}\bra{0}T\bar{\psi}(z)\Gamma\left[z,-z\right]\psi(-z)j_{\mathrm{em}}^{\sigma}(x)\ket{0}
\end{align}
where $j_{\mathrm{em}}^{\sigma}(x)=
Q\bar{\psi}(x)\gamma^{\sigma}\psi(x)$ is the electromagnetic current with an electric charge $Q$ for associated quark flavor. Contracting both sides of \eqref{eq:16} with $e^{*(\lambda)}_{\nu}$, and using \eqref{eq:11} to work out the summation over the photon polarizations,
we obtain the polarization-averaged transition matrix element between photon and vacuum, which is, in fact, the correlation function for a nonlocal quark-antiquark current and a local electromagnetic one
\begin{align}\label{eq:17}
\Pi_{\nu\Gamma}^{(\Gamma)}&=i\int\ud^4x e^{-iP\cdot x}(-g_{\nu\sigma}+\frac{P_{\nu}P_{\sigma}}{P^2})\nonumber\\
&\times\bra{0}T\bar{\psi}(z)\Gamma\left[z,-z\right]\psi(-z)j_{\mathrm{em}}^{\sigma}(x)\ket{0}
\end{align}
obeying
\begin{equation}\label{eq:18}
P^{\nu}\Pi_{\nu\Gamma}^{(\Gamma)}=0
\end{equation}
An important character of this correlation function is that it is gauge-invariant which enable us to use an appropriate form of quark propagator to evaluate it.

The explicit expressions of the correlation functions are
\begin{align}\label{eq:19}
\Pi_{\nu\mu\rho}^{(T)}
=&if^{(t)}_{\gamma\perp}(P^2)t^{(1)}_{\nu\mu\rho}\int^1_0\ud u e^{i\xi p\cdot z}\phi_{\gamma\perp}^{(t)}(u,P^2)\nonumber\\
&+if^{(t)}_{\gamma\parallel}(P^2)t^{(2)}_{\nu\mu\rho}\int^1_0\ud u e^{i\xi p\cdot z}h_{\gamma\parallel}^{(t)}(u,P^2)\nonumber\\
&+if^{(t)}_{\gamma3}(P^2)t^{(3)}_{\nu\mu\rho}\int^1_0\ud u e^{i\xi p\cdot z}h_{\gamma3}^{(t)}(u,P^2),
\end{align}
with the Lorentz structures $t^{(i)}_{\nu\mu\rho}$ being defined as
\begin{align}
&t^{(1)}_{\nu\mu\rho}=(g_{\nu\rho}p_{\mu}-g_{\nu\mu}p_{\rho})
+p_{\nu}\frac{p_{\rho}n_{\mu}-p_{\mu}n_{\rho}}{p\cdot n}\label{eq:20}\\
&t^{(2)}_{\nu\mu\rho}=(p_{\mu}n_{\rho}-p_{\rho}n_{\mu})
(p_{\nu}-\frac{P^2}{2p\cdot n}n_{\nu})
\frac{1}{p\cdot n}\label{eq:21}\\
&t^{(3)}_{\nu\mu\rho}=\frac{P^2}{2p\cdot n}[(g_{\nu\rho}n_{\mu}-g_{\nu\mu}n_{\rho})+n_{\nu}
\frac{p_{\mu}n_{\rho}-p_{\rho}n_{\mu}}{p\cdot n}]\label{eq:22}
\end{align}
for the tensor case, and
\begin{align}\label{eq:23}
\Pi_{\nu\mu}^{(V)}=&Mf^{(v)}_{\gamma\parallel}(P^2) v^{(1)}_{\nu\mu}\int^1_0\ud u e^{i\xi p\cdot z}\phi_{\gamma\parallel}^{(v)}(u,P^2)\nonumber\\
&+Mf^{(v)}_{\gamma\perp}(P^2)v^{(2)}_{\nu\mu}\int^1_0\ud u e^{i\xi p\cdot z}g_{\gamma\perp}^{(v)}(u,P^2)\nonumber\\
&-Mf^{(v)}_{\gamma3}(P^2)v^{(3)}_{\nu\mu}
\int^1_0\ud u e^{i\xi p\cdot z}g_{\gamma3}^{(v)}(u,P^2).
\end{align}
with
\begin{align}
&v^{(1)}_{\nu\mu}=\frac{p_{\mu}p_{\nu}}{P^2}-
\frac{p_{\mu}n_{\nu}}{2p\cdot n}\label{eq:24}\\
&v^{(2)}_{\nu\mu}=-g_{\mu\nu}+
\frac{p_{\mu}n_{\nu}+p_{\nu}n_{\mu}}{p\cdot n}\label{eq:25}\\
&v^{(3)}_{\nu\mu}=\frac{n_{\mu}}{2p\cdot n}(p_{\nu}-\frac{P^2}{2p\cdot n}n_{\nu})\label{eq:26}
\end{align}
for the vector case, and
\begin{align}\label{eq:27}
\Pi_{\nu}^{(S)}=if^{(s)}_{\gamma\parallel}(P^2)\left[(p\cdot z)p_{\nu}-\frac{1}{2}z_{\nu}P^2\right]
\int^1_0\ud u e^{i\xi p\cdot z}h^{(s)}_{\gamma\parallel}(u,P^2)
\end{align}
for the scalar case, and
\begin{align}\label{eq:28}
\Pi_{\nu\mu}^{(A)}=M  f^{(a)}_{\gamma\perp}(P^2)\epsilon_{\mu\nu\alpha\beta}p^{\alpha}z^{\beta}\int^1_0\ud u e^{i\xi p\cdot z}g^{(a)}_{\gamma\perp}(u,P^2),
\end{align}
for the axial vector case, respectively.

Now we extract an individual photon DA from the corresponding correlation function $\Pi_{\nu\Gamma}^{(\Gamma)}$ by contracting it with an appropriate projection operator constructed from $p_{\mu}$, $z_{\mu}$ and $g_{\mu\nu}$, and then performing the inverse Fourier transform
\begin{align}\label{eq:29}
F[\phi^{(i)}_{\gamma}(\tau,P^2)]
\equiv \frac{p^+}{\pi}\int d\tau e^{i2u p^+\tau}\phi^{(i)}_{\gamma}(\tau,P^2)
=\phi^{(i)}_{\gamma}(u,P^2).
\end{align}
At the end, all photon DAs could be expressed in terms of correlation functions as
\begin{align}\label{eq:30}
f^{(i)}_{\gamma}(P^2)\phi^{(i)}_{\gamma}(u,P^2)
=F\left[-it_{(i)}^{\nu\dots}\Pi_{\nu\Gamma}^{(\Gamma)}\right].
\end{align}
The details of the calculation are given out in Appendix.\ref{app:aqqA}.

After working out the correlation functions, the series of \eqref{eq:30} can be used to determine the various coupling constants
$f_i(P^2)$ by integrating both sides of these equations over $u$ from $0$ to $1$, and the various photon DAs $\phi^{(i)}_{\gamma}(u,P^2)$ can be determined by substituting the corresponding coupling constants $f_i(P^2)$ into the equations. We note here that the solutions we have obtained are universal in the sense that their validity is independent of the special dynamical model adopted in calculation of the correlation function.

\section{Spectral representations of correlation functions}
To obtain the explicit expressions of photon DAs from \eqref{eq:30}, we need to choose a dynamical model to calculate the correlation functions. Since we are interesting in the low-energy region, we would like to work in the low-energy effective theory based on instanton vacuum model of QCD, in which the essential element in our calculation is the effective quark propagator.

The effective quark propagator in the instanton vacuum model has been derived in the singular gauge of  instantons\cite{qpropagator} by applying
the Feynman variational principle\cite{FeynVP} to the low-energy effective theory of QCD. The effective chiral action which describes the interaction of quarks
with an external meson field $U$ in the large $N_c$ limit is
\begin{equation}\label{eq:31}
S_{eff}=-N_c\mathrm{tr}\ln(i\slashed{\partial}+iMFUF),
\end{equation}
where
\begin{equation}
U=\exp[i\gamma_5\tau^a\pi^a(x)]
\label{eq:32}
\end{equation}
and $M$ is the dynamical quark mass at zero momentum. $F(k)$ is the form factor, related to the Fourier transform of the instanton zero mode
\begin{equation}\label{eq:33}
F(k)=2y[I_0(y)K_1(y)-I_1(y)K_0(y)]-2I_1(y)K_1(y)
\end{equation}
with $y=k\rho/2$ and $\rho=(600\mathrm{MeV})^{-1}$ being the typical inverse instanton size. In this low-energy effective theory, the various correlation functions can be evaluated by a quark-loop in the large $N_c$ limit with the following effective quark propagator
\begin{equation}\label{eq:34}
S_F(k)=\frac{\slashed{k}+MF^2(k)}{k^2-M^2F^4(k)+i\epsilon}.
\end{equation}

For the sake of convenience, we choose to work with the pole form of the quark form factor $F(k)$\cite{prd1999pi}
\begin{equation}
F(k)=\left(\frac{-\Lambda^2}{k^2-\Lambda^2+i\epsilon}\right)^n
\label{eq:35}
\end{equation}
where $\Lambda$ and $n$ are the artificial but justified input parameters.

To check whether the propagator \eqref{eq:34} is theoretically acceptable, and to express the correlation functions in some more general form, we write \eqref{eq:34} in terms of the spectral densities $\rho_1(\omega^2)$ and $\rho_2(\omega^2)$ defined in a general formula for fermion
propagator\cite{Itzykson}
\begin{equation}\label{eq:36}
S_F(k)=\int \ud \omega^2\frac{\rho_1(\omega^2)\slashed{k}+\rho_2(\omega^2)}
{k^2-\omega^2+i\epsilon}
\end{equation}
Comparing \eqref{eq:34} and \eqref{eq:36}, we have
\begin{align}
\int \ud \omega^2\frac{\rho_1(\omega^2)}{k^2-\omega^2+i\epsilon}&=
\frac{(k^2-\Lambda^2)^{4n}}{k^2(k^2-\Lambda^2)^{4n}-(M \Lambda^{4n})^2}\label{eq:37},\\
\int \ud \omega^2\frac{\rho_2(\omega^2)}{k^2-\omega^2+i\epsilon}&=
\frac{M\Lambda^{4n}(k^2-\Lambda^2)^{2n}}
{k^2(k^2-\Lambda^2)^{4n}-(M \Lambda^{4n})^2}\label{eq:38}.
\end{align}
After making the expansion in simple partial fractions
\begin{align}
\frac{(k^2-\Lambda^2)^{4n}}{k^2(k^2-\Lambda^2)^{4n}-(M \Lambda^{4n})^2}&=\sum_{i=1}^{4n+1}
\frac{f_i(z_i-\Lambda^2)^{4n}}{k^2-z_i}\label{eq:39},\\
\frac{M\Lambda^{4n}(k^2-\Lambda^2)^{2n}}
{k^2(k^2-\Lambda^2)^{4n}-(M \Lambda^{4n})^2}&=M\Lambda^{4n}\sum_{i=1}^{4n+1}
\frac{f_i(z_i-\Lambda^2)^{2n}}{k^2-z_i}\label{eq:40},
\end{align}
where $z_i$ are the roots of an algebraic equation
\begin{equation}\label{eq:41}
z(z-\Lambda^2)^{4n}=(M\Lambda^{4n})^2
\end{equation}
and
$$f_i=\prod^{4n+1}_{j=1,j\neq i}\frac{1}{z_j-z_i},$$
the spectral densities can be easily read off as the weighted
summations of Dirac $\delta$-functions of $\omega^2$ peaked at different roots $z_i$
\begin{align}
\rho_1(\omega^2)&=\sum_{i=1}^{4n+1} f_i(z_i-\Lambda^2)^{4n}\delta(\omega^2-z_i)\nonumber\\
&=\delta\left[\prod_i^{4n+1}\frac{\omega^2-z_i}{(\omega^2+\Lambda^2)^{4n}}\right]\label{eq:42},\\
\rho_2(\omega^2)&=M\Lambda^{4n}\sum_{i=1}^{4n+1}f_i(z_i-\Lambda^2)^{2n}\delta(\omega^2-z_i)\nonumber\\
&=M\Lambda^{4n}\delta\left[\prod_i^{4n+1}\frac{\omega^2-z_i}{(\omega^2+\Lambda^2)^{2n}}\right]\label{eq:43}.
\end{align}
which satisfy the three constraints for the fermion's spectral densities
\begin{align}
&(\mathrm{\romannumeral1})\quad \rho_1(\omega^2)\quad\mathrm{and}\quad\rho_2(\omega^2)\quad\mathrm{are}\quad\mathrm{real},\nonumber\\
&(\mathrm{\romannumeral2})\quad \rho_1(\omega^2)\ge0,\nonumber\\
&(\mathrm{\romannumeral3})\quad \omega\rho_1(\omega^2)-\rho_2(\omega^2)\ge0.\nonumber
\end{align}
This reveals some consistency of the form of the effective quark propagator \eqref{eq:34} with $F(k)$ defined by \eqref{eq:35}.

\begin{figure*}
\begin{minipage}{0.45\linewidth}
\centerline{\includegraphics{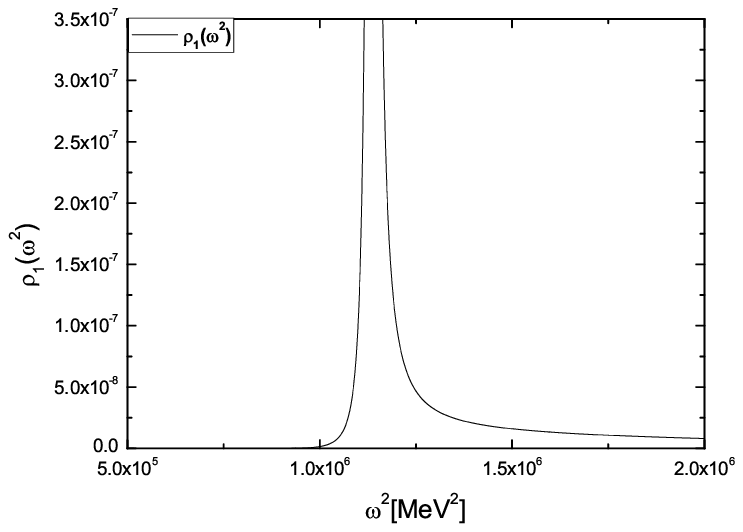}}
\end{minipage}
\hfill
\begin{minipage}{0.45\linewidth}
\centerline{\includegraphics{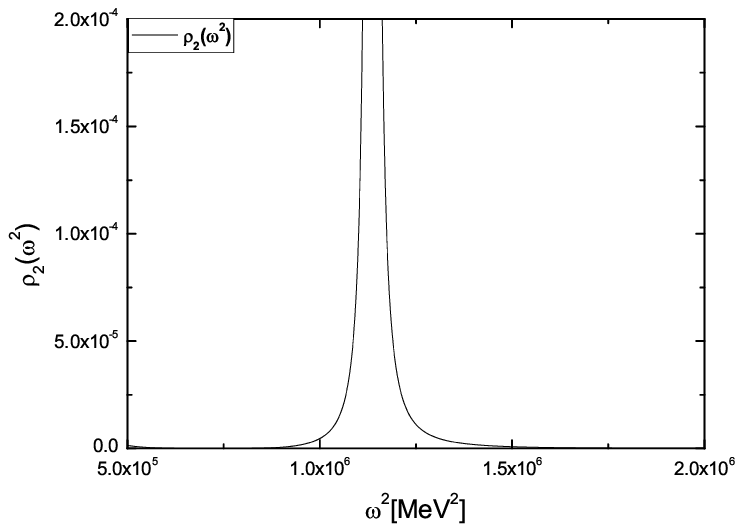}}
\end{minipage}
\begin{minipage}{0.45\linewidth}
\centerline{\includegraphics{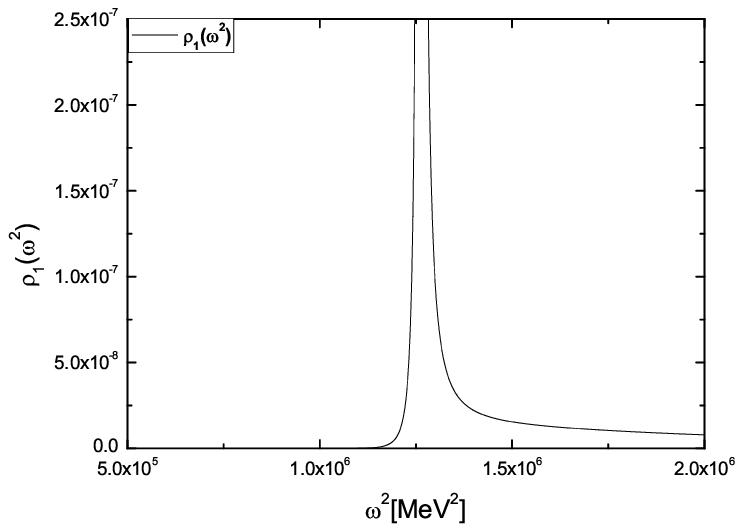}}
\end{minipage}
\hfill
\begin{minipage}{0.45\linewidth}
\centerline{\includegraphics{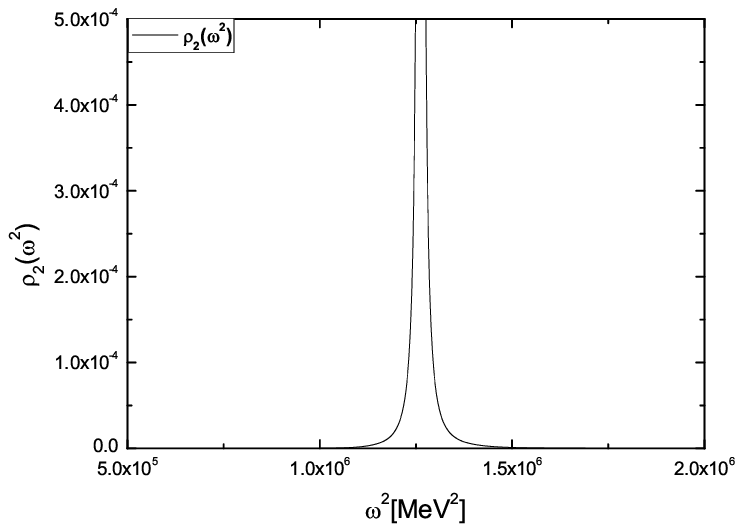}}
\end{minipage}
\caption{\label{fig:spec}The spectral density $\rho_{1}(\omega^2)$ and $\rho_{2}(\omega^2)$  for $n=1$(upper case) and $n=2$(lower case) and $\Lambda=850\mathrm{MeV}$.}
\end{figure*}

The shape of spectral densities are displayed in Fig.\ref{fig:spec}, where the different roots $z_i$ are shown to be in between $1\mathrm{GeV}^2$ and $1.5\mathrm{GeV}^2$, which is just the very energy regime of effective low-energy theory of QCD. It can be seen from the mentioned figures that the spectral functions $\rho_1$ and $\rho_2$ are almost coincident for the cases of $n=1$ and $n=2$. This fact is in agreement with the statement that the forms of the photon DAs are insensitive respect to a change of value of the artificial variable $n$\cite{yuran06prd}. For this reason, it may be safely for us to choose, say $n=1$ and $\Lambda=850\mathrm{MeV}$, in our numerical simulation.

Using the Wick's theorem and the general spectral form of quark propagator, the correlation functions for various currents can be expressed analytically as follows
\begin{align}
\Pi_{\nu\mu\rho}^{(T)}&=C_{\nu\sigma}\hat{I}
\left[g^{\sigma}_{\rho}[(P-k)_{\mu}F_1+k_{\mu}F_2]\right.\nonumber\\
&\left.-g^{\sigma}_{\mu}[(P-k)_{\rho}F_1+k_{\rho}F_2]\right]\label{eq:44},\\
\Pi_{\nu\mu}^{(V)}&=C_{\nu\sigma}\hat{I}
\left[F_3[k^{\sigma}(P-k)_{\mu}+k_{\mu}(P-k)^{\sigma}]\right.\nonumber\\
&\left.-g^{\sigma}_{\mu}[F_3(P-k)\cdot k+F_4]\right]\label{eq:45},\\
\Pi_{\nu}^{(S)}&=C_{\nu\sigma}\hat{I}
\left[(k^{\sigma}-P^{\sigma})F_1
+k^{\sigma}F_2\right]\label{eq:46},\\
\Pi_{\nu\mu}^{(A)}&=C_{\nu\sigma}\hat{I}
F_3{\epsilon^{\sigma}}_{\mu\rho\alpha}
P^{\rho}k^{\alpha}\label{eq:47}.
\end{align}
where $C_{\nu\sigma}$ and $\hat{I}$ are a constant tensor and an operator respectively defined by
\begin{align*}
&C_{\nu\sigma}\equiv-4QN_c(-g_{\nu\sigma}
+\frac{P_{\nu}P_{\sigma}}{P^2}),\\
&\hat{I}\equiv
\int\frac{\ud^4k}{(2\pi)^4} e^{-i(p-2k)\cdot z}\int\frac{\ud\omega}{(P-k)^2-\omega}\int\frac{\ud\mu}{k^2-\mu},
\end{align*}
and $F_i$ the products of the two appropriate spectral densities $\rho_j(\omega)$ and $\rho_k(\mu)$ of the effective quark propagator
\begin{align*}
&F_1\equiv\rho_1(\omega)\rho_2(\mu),\quad F_2\equiv\rho_1(\mu)\rho_2(\omega),\\
&F_3\equiv\rho_1(\omega)\rho_1(\mu),\quad F_4\equiv\rho_2(\omega)\rho_2(\mu).
\end{align*}

\section{Explicit analytic expressions of photon DAs}\label{sec5}
\subsection{chiral odd}

Now, we are in a position to be able to express the light-cone photon DAs in terms of the functions $F_i$. Substituting \eqref{eq:44} and \eqref{eq:46} into \eqref{eq:a12}-\eqref{eq:a14} and \eqref{eq:a18}, and carrying out the inverse Fourier transformation, we obtain the explicit expressions of the chiral-odd light-cone photon DAs
\begin{align}
\phi^{(t)}_{\gamma\perp}(u,P^2)&=\frac{4N_c}{if^{(t)}_{\gamma\perp}(P^2)}\hat{D}F^{(1)}          \label{eq:48},\\
h^{(t)}_{\gamma\parallel}(u,P^2)&=\frac{4N_c}{if^{(t)}_{\gamma\parallel}(P^2)}\hat{D}F^{(2)}     \label{eq:49},\\
h^{(t)}_{\gamma3}(u,P^2)&=\frac{4N_c}{if^{(t)}_{\gamma3}(P^2)}\hat{D}F^{(3)}                     \label{eq:50},\\
h^{(s)}_{\gamma\parallel}(u,P^2)&=\frac{4N_c}{if^{(s)}_{\gamma\parallel}(P^2)}\hat{T}F^{(4)}      \label{eq:51},
\end{align}
where the operations $\hat{D}$ and $\hat{T}$ are defined as the following three-fold integrations
\begin{eqnarray*}
\hat{D}\equiv\int\frac{\ud^4k}{(2\pi)^4}\delta(k^+-up^+)\int\frac{\ud\omega}{(P-k)^2-\omega}\int\frac{\ud\mu}{k^2-\mu},\\
\hat{T}\equiv\int\frac{\ud^4k}{(2\pi)^4}\theta(k^+-up^+)\int\frac{\ud\omega}{(P-k)^2-\omega}\int\frac{\ud\mu}{k^2-\mu},
\end{eqnarray*}
and $F^{(k)}$ for $k$ from 1 to 4 are listed in Table \ref{tab:fv} at the end of this section.

Integrating both sides of \eqref{eq:48}-\eqref{eq:51} over $u$ from $0$ to $1$, and using the normalization conditions of the light-cone DAs, the corresponding couplings, $f^{(t)}_{\gamma\perp}$, $f^{(t)}_{\gamma\parallel}$, $f^{(t)}_{\gamma3}$ and $f^{(s)}_{\gamma\parallel}$, are obtained to be
\begin{align}
f^{(t)}_{\gamma\perp}(P^2)
&=-4iN_c\int_0^1\ud u\hat{D}F^{(1)}\label{eq:52},\\
f^{(t)}_{\gamma\parallel}(P^2)
&=-4iN_c\int_0^1\ud u\hat{D}F^{(2)}\label{eq:53},\\
f^{(t)}_{\gamma3}(P^2)
&=-4iN_c\int_0^1\ud u\hat{D}F^{(3)}\label{eq:54},\\
f^{(s)}_{\gamma\parallel}(P^2)
&=-4iN_c\int_0^1\ud u\hat{T}F^{(4)}\label{eq:55}.
\end{align}

To obtain the explicit expressions of the DAs, we need to evaluate the integrations of forms like
\begin{align}
I_1&=\int\frac{\ud^4k}{(2\pi)^4}\frac{\delta(k^+-up^+)}
{[(P-k)^2-\omega](k^2-\mu)}\label{eq:56},\\
I_2&=\int\frac{\ud^4k}{(2\pi)^4}\frac{(k\cdot p)\delta(k^+-up^+)}
{[(P-k)^2-\omega](k^2-\mu)}\label{eq:57}.
\end{align}
Let us focus on the general Lorentz structure of $I_2$, an integration over the integrand involving factor of $k_{\mu}$,  which should be of the form after Lorentz decomposition
\begin{equation}\label{eq:58}
\int\frac{\ud^4k}{(2\pi)^4}\frac{\delta(k^+-up^+)k^{\mu}}
{[(P-k)^2-\omega](k^2-\mu)}=A(u)P^{\mu}+B(u)n^{\mu},
\end{equation}
where $A(u)$ and $B(u)$ are some scalar functions, and there is an obvious condition for $B(u)$,
\begin{equation}\label{eq:59}
\int_0^1du B(u)=0,
\end{equation}
due to Lorentz coverance. This property will be useful when we calculate the coupling constants by means of the normalization conditions.

Turn to evaluate $I_1$ and $I_2$. Introducing the dimensionless variables $\eta=p^+k^-/\Lambda^2$, $s=P^2/\Lambda^2$, and $t=|k_{\perp}|^2/\Lambda^2$, and completing the integration over $k^+$ and $\eta$, the integrals $I_1$ and $I_2$ can be worked out to be (see Appendix \ref{app:appB})
\begin{align}
I_1&=\frac{i}{2(2\pi)^2}\ln(1+\frac{v}{-u\bar{u}s+u\omega
+\bar{u}\mu})\label{eq:60},\\
I_2&=-\frac{i}{4(2\pi)^2}(\omega-\mu-\bar{u}s)
\ln(1+\frac{v}{-u\bar{u}s+u\omega+\bar{u}\mu})\label{eq:61}.
\end{align}
where $v$ is a cutoff for the upper bound of the absolute value of $k$.

Now, using the formulae \eqref{eq:60} and \eqref{eq:61}, we can carry out the integration over $k$, and then the integration over $\mu$ and $\omega$,
and finally obtain the explicit expressions of the light-cone photon DAs for tensor and scalar cases
\begin{align}
&\phi^{(t)}_{\gamma\perp}(u,P^2)=
-\frac{2N_cM\Lambda^4}{f^{(t)}_{\gamma\perp}(P^2)(2\pi)^2}
\sum^5_{i,j=1}W_{ij}V^{(1)}_{ij}\label{eq:62},\\
&h^{(t)}_{\gamma\parallel}(u,P^2)=
-\frac{2N_cM\Lambda^4}{f^{(t)}_{\gamma\parallel}(P^2)(2\pi)^2}
\sum^5_{i,j=1}W_{ij}V^{(2)}_{ij}\label{eq:63},\\
&h^{(t)}_{\gamma3}(u,P^2)=
-\frac{2N_cM\Lambda^4}{f^{(t)}_{\gamma3}(P^2)(2\pi)^2}
\sum^5_{i,j=1}W_{ij}V^{(3)}_{ij}\label{eq:64},\\
&h^{(s)}_{\gamma\parallel}(u,P^2)=
-\frac{2N_cM\Lambda^4}{f^{(s)}_{\gamma\parallel}(P^2)(2\pi)^2}
\sum^5_{i,j=1}W_{ij}V^{(4)}_{ij}\label{eq:65},
\end{align}
where
\begin{displaymath}
W_{ij}=f_if_j(z_i-\Lambda^2)^4(z_j-\Lambda^2)^4
\ln(-u\bar{u}s+uz_i+\bar{u}z_j).
\end{displaymath}
and the explicit expressions of $V^{(k)}_{ij}$ are listed in Table \ref{tab:fv}.

In the above expressions, the cutoff $v$ disappears because of the identity for the partial fraction summation
\begin{equation}\label{eq:66}
\sum^5_{i=1}f_iz^n_i=0\quad\mathrm{for}\quad n<4.
\end{equation}

We note here that in evaluating the light-cone photon DA corresponding to the scalar current, we need to deal with the integrals
\begin{align*}
I_3&=\int\frac{\ud^4k}{(2\pi)^4}\frac{\theta(k^+-up^+)}
{[(P-k)^2-\omega](k^2-\mu)},\\
I_4&=\int\frac{\ud^4k}{(2\pi)^4}\frac{(k\cdot p)\theta(k^+-up^+)}
{[(P-k)^2-\omega](k^2-\mu)}.
\end{align*}
which are the integrations of $I_1$ and $I_2$ over $u$, respectively. Therefore, only $\partial h^{(s)}_{\gamma\parallel}/\partial u$ (not $h^{(s)}_{\gamma\parallel}$) can be evaluated in the same way as
the photon DAs corresponding to the tensor current. To obtain $h^{(s)}_{\gamma\parallel}$, we need a boundary conditions at $u=0$ and $u=1$, which, for the sake of simplicity, are assumed to be
\begin{equation}\label{eq:67}
h^{(s)}_{\gamma\parallel}(0,P^2)=h^{(s)}_{\gamma\parallel}(1,P^2)=0
\end{equation}
The same is done for the light-cone photon DA $g^{(a)}_{\gamma\perp}$ corresponding to the axial current, i.e.,
\begin{equation}\label{eq:68}
g^{(a)}_{\gamma\perp}(0,P^2)=g^{(a)}_{\gamma\perp}(1,P^2)=0
\end{equation}

Applying the normalization conditions \eqref{eq:2}, the corresponding coupling constants are obtained to be
\begin{align}
&f^{(t)}_{\gamma\perp}(P^2)=
-\frac{2N_cM\Lambda^4}{(2\pi)^2}\sum^5_{i,j=1}\int_0^1\ud u W_{ij} V^{(1)}_{ij}\label{eq:69},\\
&f^{(t)}_{\gamma\parallel}(P^2)=
-\frac{2N_cM\Lambda^4}{(2\pi)^2}\sum^5_{i,j=1}\int_0^1\ud u W_{ij} V^{(2)}_{ij}\label{eq:70},\\
&f^{(t)}_{\gamma3}(P^2)=
-\frac{2N_cM\Lambda^4}{(2\pi)^2}\sum^5_{i,j=1}\int_0^1\ud u W_{ij} V^{(3)}_{ij}\label{eq:71},\\
&f^{(s)}_{\gamma\parallel}(P^2)=
-\frac{2N_cM\Lambda^4}{(2\pi)^2}\sum^5_{i,j=1}\int_0^1\ud u W_{ij} V^{(4)}_{ij}\label{eq:72},
\end{align}
We note here that in obtaining the results of \eqref{eq:69}-\eqref{eq:72} (similarly for \eqref{eq:85}-\eqref{eq:88}), we have used the fact that the integration of terms containing a factor of $k^-$ over $u$ is vanishing, as expected from the identity \eqref{eq:59}.

\subsection{chiral even}
For the chiral-even DAs, substituting \eqref{eq:45} and \eqref{eq:47} into \eqref{eq:a15}-\eqref{eq:a17} and \eqref{eq:a19} respectively, we have
\begin{align}
\phi^{(v)}_{\gamma\parallel}(u,P^2)
&=\frac{4iN_c}{M f^{(v)}_{\gamma\parallel}(P^2)}\hat{D} F^{(5)}\label{eq:73},\\
g^{(v)}_{\gamma\perp}(u,P^2)
&=\frac{4iN_c}{M f^{(v)}_{\gamma\perp}(P^2)}\hat{D}F^{(6)}\label{eq:74},\\
g^{(v)}_{\gamma3}(u,P^2)
&=\frac{4iN_c}{M f^{(v)}_{\gamma3}(P^2)}\hat{D}F^{(7)}\label{eq:75},\\
g^{(a)}_{\gamma\perp}(u,P^2)
&=\frac{4iN_c}{M  f^{(a)}_{\gamma\perp}(P^2)}\hat{T}F^{(8)}\label{eq:76}.
\end{align}

The corresponding coupling constants are
\begin{align}
f^{(v)}_{\gamma\parallel}(P^2)
&=\frac{4iN_c}{M}\int_0^1\ud u\hat{D}
F^{(5)}\label{eq:77},\\
f^{(v)}_{\gamma\perp}(P^2)
&=\frac{4iN_c}{M}\int_0^1\ud u\hat{D}
F^{(6)}\label{eq:78},\\
f^{(v)}_{\gamma3}(P^2)
&=\frac{4iN_c}{M}\int_0^1\ud u\hat{D}
F^{(7)}\label{eq:79},\\
f^{(a)}_{\gamma\perp}(P^2)
&=\frac{4iN_c}{M}\int_0^1\ud u\hat{T}
F^{(8)}\label{eq:80}.
\end{align}

After completing the integrations over $k$, $\mu$ and $\omega$ just as the same way as in the chiral-odd case, we have the explicit expressions
\begin{align}
\phi^{(v)}_{\gamma\parallel}(u,P^2)&=
\frac{2N_c}{M f^{(v)}_{\gamma\parallel}(P^2)(2\pi)^2}
\sum^5_{i,j=1}W_{ij}V^{(5)}_{ij}\label{eq:81},\\
g^{(v)}_{\gamma\perp}(u,P^2)&=
\frac{2N_c}{M f^{(v)}_{\gamma\perp}(P^2)(2\pi)^2}
\sum^5_{i,j=1}W_{ij}V^{(6)}_{ij}\label{eq:82},\\
g^{(v)}_{\gamma3}(u,P^2)&=
\frac{2N_c}{M f^{(v)}_{\gamma3}(P^2)(2\pi)^2}
\sum^5_{i,j=1}W_{ij}V^{(7)}_{ij}\label{eq:83},\\
g^{(a)}_{\gamma\perp}(u,P^2)&=
\frac{2N_c\Lambda^2}{M f^{(a)}_{\gamma\perp}(P^2)(2\pi)^2}
\sum^5_{i,j=1}W_{ij}V^{(8)}_{ij}\label{eq:84}.
\end{align}
for the chiral-even light-cone photon DAs, and
\begin{align}
f^{(v)}_{\gamma\parallel}(P^2)&=\frac{2N_c}{M(2\pi)^2}
\sum^5_{i,j=1}W_{ij}\int_0^1\ud u V^{(5)}_{ij}\label{eq:85},\\
f^{(v)}_{\gamma\perp}(P^2)&=\frac{2N_c}{M(2\pi)^2}
\sum^5_{i,j=1}W_{ij}\int_0^1\ud u V^{(6)}_{ij}\label{eq:86},\\
f^{(v)}_{\gamma3}(P^2)&=\frac{2N_c}{M(2\pi)^2}
\sum^5_{i,j=1}W_{ij}\int_0^1\ud u V^{(7)}_{ij}\label{eq:87},\\
f^{(a)}_{\gamma\perp}(P^2)&=\frac{2N_c\Lambda^2}{M (2\pi)^2}
\sum^5_{i,j=1}W_{ij}\int_0^1\ud u V^{(8)}_{ij}\label{eq:88}.
\end{align}
for the corresponding coupling constants, where
\begin{displaymath}
M_{ij}=(M\Lambda^4)^2(z_j-\Lambda^2)^{-2}(z_i-\Lambda^2)^{-2}.
\end{displaymath}
and the $F^{(k)}$ and $V_{ij}^{(k)}$ for $k$ from 5 to 8 are listed in Table \ref{tab:fv}.

\begin{table*}[!h]
\caption{The explicit expressions of $F^{(k)}$ and $V_{ij}^{(k)}$\label{tab:fv}}
\center
\begin{tabular}{lll}\hline
$k$         & $F^{(k)}$ & $V_{ij}^{(k)}$  \\ \hline
$1$ & $F_1+\frac{k^+}{P^+}(F_2-F_1)$ & $\bar{u}(z_j-\Lambda^2)^{-2}+u(z_i-\Lambda^2)^{-2}$  \\
$2$ & $F_1+(\frac{k^+}{2P^+}+\frac{k\cdot p}{P^2})(F_2-F_1)$ & $(s+z_i-z_j)(z_j-\Lambda^2)^{-2}+(s+z_j-z_i)(z_i-\Lambda^2)^{-2}$ \\
$3$ & $F_1+\frac{2k\cdot p}{P^2}(F_2-F_1)$ & $(us+z_i-z_j)(z_j-\Lambda^2)^{-2}+(\bar{u}s+z_i-z_j)(z_i-\Lambda^2)^{-2}$ \\
$4$ & $(\frac{k\cdot z}{p\cdot z}-\frac{2k\cdot p}{P^2})(F_1+F_2)$ & $\left[(z_j-\Lambda^2)^{-2}+(z_i-\Lambda^2)^{-2}\right](u\bar{u}s-uz_i-\bar{u}z_j)$ \\
$5$ & $F^{(6)}-(1-\frac{2k\cdot z}{p\cdot z})F^{(8)}$ & $M_{ij}+\bar{u}z_j+uz_i-2u\bar{u}s$  \\
$6$ & $F_3(P-k)\cdot k+F_4$ & $M_{ij}+s-(z_i+z_j)$  \\
$7$ & $F^{(6)}+(1-\frac{4k\cdot p}{P^2})F^{(8)}$ & $M_{ij}-2u\bar{u}s+(3uz_i-3uz_j+2z_j-z_i)$  \\
$8$ & $[k\cdot p-\frac{k\cdot z}{2p\cdot z}P^2]F_3$ & $-u\bar{u}s+uz_i+\bar{u}z_j$  \\ \hline
\end{tabular}
\end{table*}

\section{numerical result}\label{sec6}
The input parameters in our numerical simulation are as follows: The color number $N_c$ is taken to be three; The parameter $n$ in the pole form of the effective quark propagator is chosen to be one with $\Lambda=850\mathrm{MeV}$, (We have demonstrated that the spectral densities of the effective quark propagator is almost independent of a change of $n$); The mass scale $M$ in the pole-form effective quark propagator is taken to be $M=350\mathrm{MeV}$.

\begin{figure*}
\begin{minipage}{0.45\linewidth}
\centerline{\includegraphics{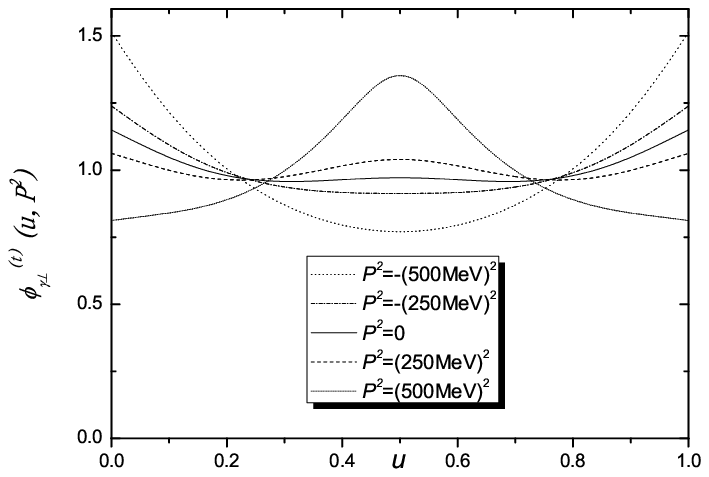}}
\end{minipage}
\begin{minipage}{0.45\linewidth}
\centerline{\includegraphics{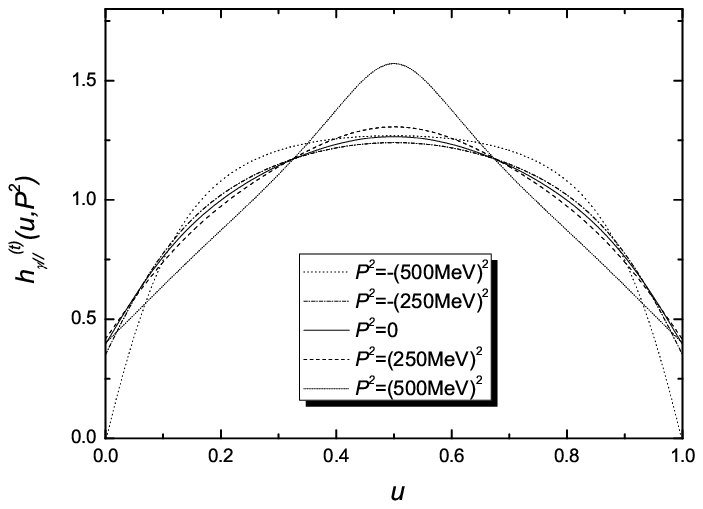}}
\end{minipage}
\begin{minipage}{0.45\linewidth}
\centerline{\includegraphics{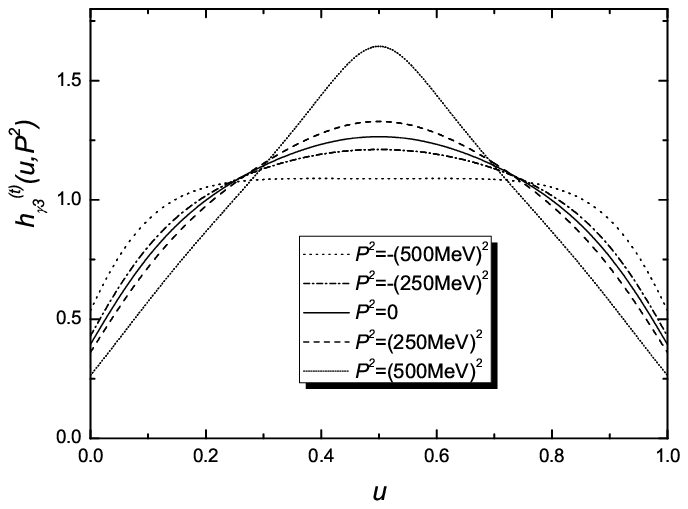}}
\end{minipage}
\hfill
\begin{minipage}{0.45\linewidth}
\centerline{\includegraphics{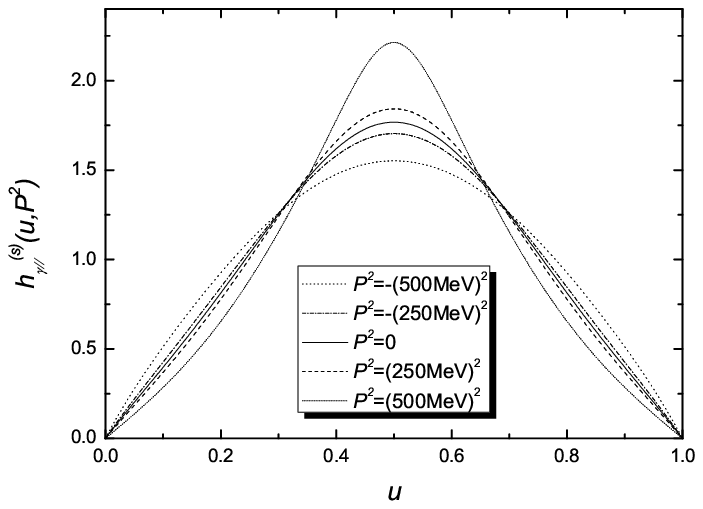}}
\end{minipage}
\caption{\label{fig:odd}The chiral-odd photon distribution amplitude for n=1 and $\Lambda=850\mathrm{MeV}$.}
\end{figure*}

\begin{figure*}
\begin{minipage}{0.45\linewidth}
\centerline{\includegraphics{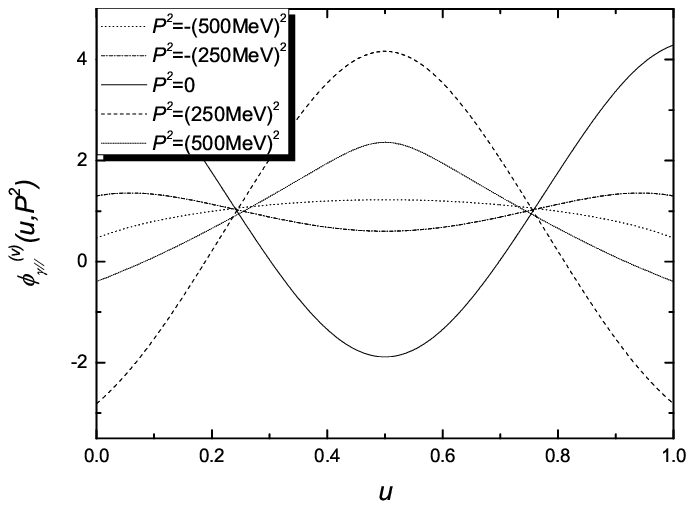}}
\end{minipage}
\begin{minipage}{0.45\linewidth}
\centerline{\includegraphics{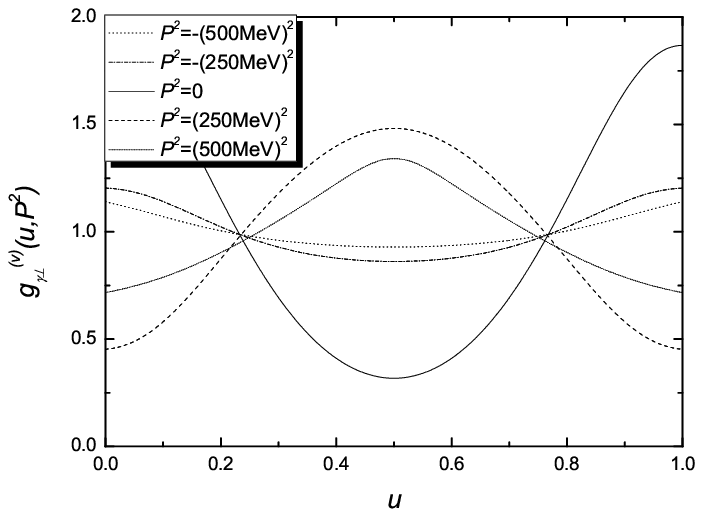}}
\end{minipage}
\begin{minipage}{0.45\linewidth}
\centerline{\includegraphics{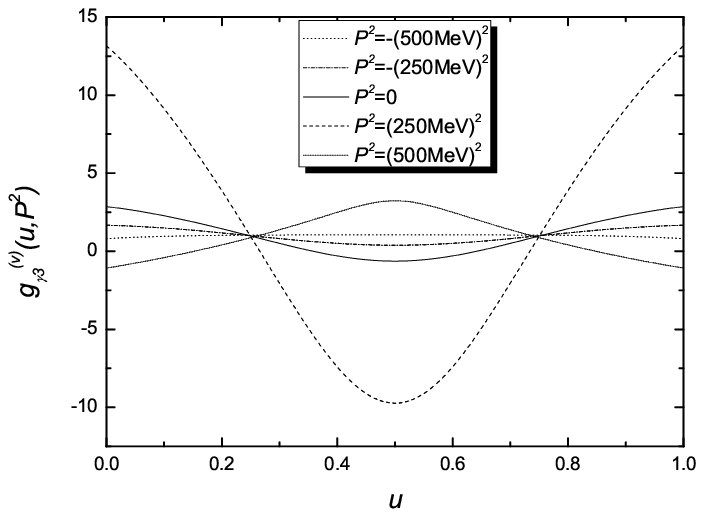}}
\end{minipage}
\hfill
\begin{minipage}{0.45\linewidth}
\centerline{\includegraphics{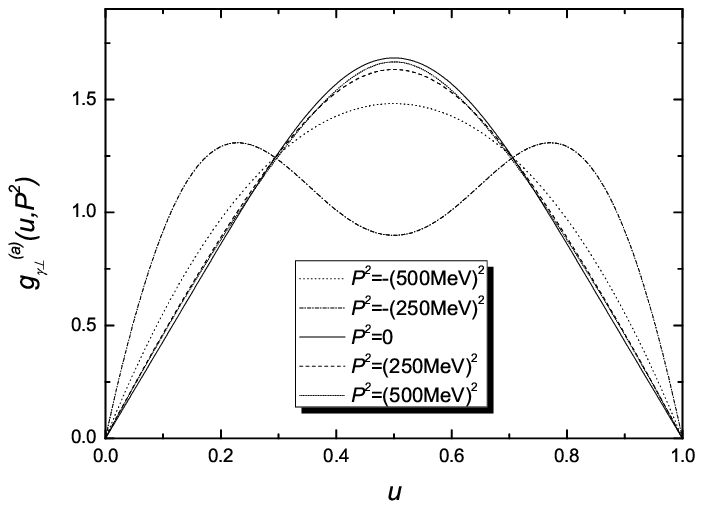}}
\end{minipage}
\caption{\label{fig:even}The chiral-even photon distribution amplitude for n=1 and $\Lambda=850\mathrm{MeV}$.}
\end{figure*}

The dependencies of the eight light-cone photon DAs on the momentum fraction $u$ for different virtuality $P^2$ are displayed respectively in Fig.\ref{fig:odd} and Fig.\ref{fig:even}, where the dot lines correspond to the virtuality of $P^2=-(500\mathrm{MeV})^2$, the dash-dot lines to $P^2=-(250\mathrm{MeV})^2$, the solid lines to $P^2=0$, the dash lines to $P^2=(250\mathrm{MeV})^2$ and short-dot lines to $P^2=(500\mathrm{MeV})^2$.

All the photon DAs are invariant under the exchange between $u$ and $\bar{u}$, and have an maximum(or minimum) at the middle, which is just what we expected from the symmetry between quark and antiquark fields in the associated currents.

Now consider the endpoint behavior of the photon DAs. From the mentioned figures, we can see that six light-cone photon DAs, namely $\phi^{(t)}_{\gamma\perp}$,
$h^{(t)}_{\gamma\parallel}$  and $h^{(t)}_{\gamma3}$ for the tensor case and $\phi^{(v)}_{\gamma\parallel}$, $g^{(v)}_{\gamma\perp}$ and
$g^{(v)}_{\gamma3}$ for the vector case, are non-vanishing at the endpoints $u=0$ and $u=1$, whereas the other two DAs, namely
$h^{(s)}_{\gamma\parallel}$ for the scalar case and $g^{(a)}_{\gamma\perp}$ for the axial vector case, are vanishing at the endpoints being analogous to the asymptotic light-cone pion wave functions. We note here that the end-point behavior for the latter is in fact arising from our artificial assumption for the boundary
conditions of the photon DAs corresponding to the scalar and axial currents. There may be appearance of the non-zero values at the end points for some physical arguments, say by comparison with the experimental data.

For the former, the non-vanishing end-point behaviors are still, somewhat, different for different cases. In the tensor case, $h^{(t)}_{\gamma\parallel}$ and
$h^{(t)}_{\gamma3}$ are bent down  when closer to the endpoints, and suppressed obviously at the endpoints, and there is only one single extremum at
$u=1/2$, which means that the momentum tends to be distributed equally to the quark and anti-quark. However, in the vector case, the curves of the
light-cone photon DAs, $\phi^{(v)}_{\gamma\parallel}$, $g^{(v)}_{\gamma\perp}$ and $g^{(v)}_{\gamma3}$, are gradually bent up when closer to the
endpoints. In addition, their concavity changes with varying $P^2$, and $\phi^{(v)}_{\gamma\parallel}$ and $g^{(v)}_{\gamma\perp}$
changes sign at some time-like momentum about $P^2\sim(300\mathrm{MeV})^2$ due to the normalization.

\begin{figure*}
\begin{minipage}{0.45\linewidth}
\centerline{\includegraphics{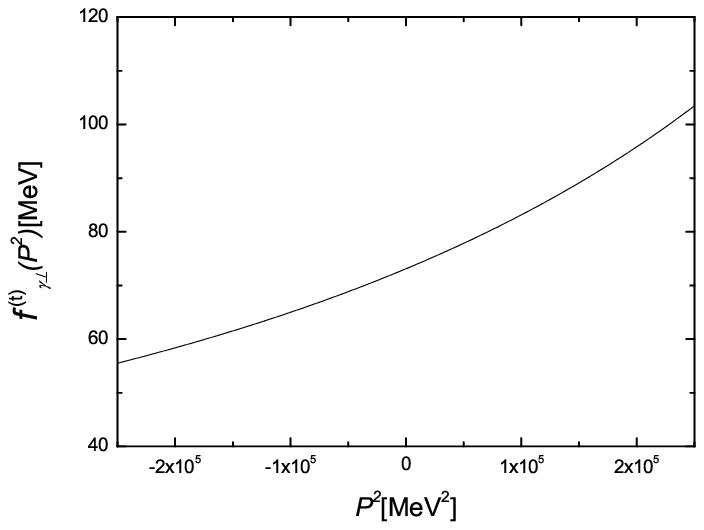}}
\end{minipage}
\begin{minipage}{0.45\linewidth}
\centerline{\includegraphics{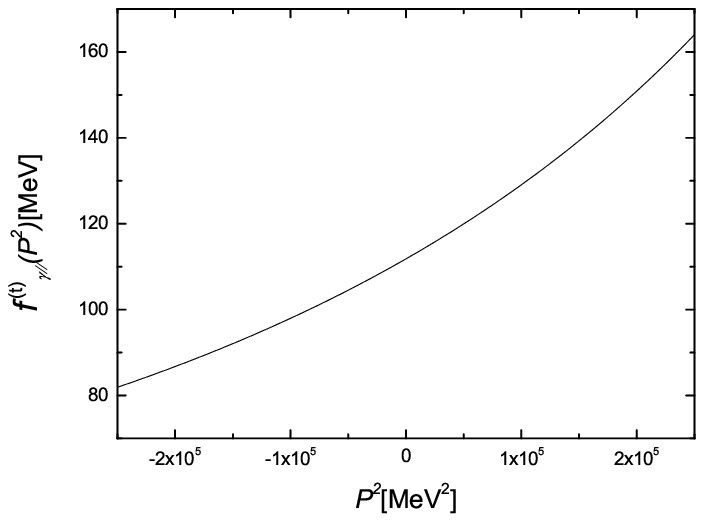}}
\end{minipage}
\begin{minipage}{0.45\linewidth}
\centerline{\includegraphics{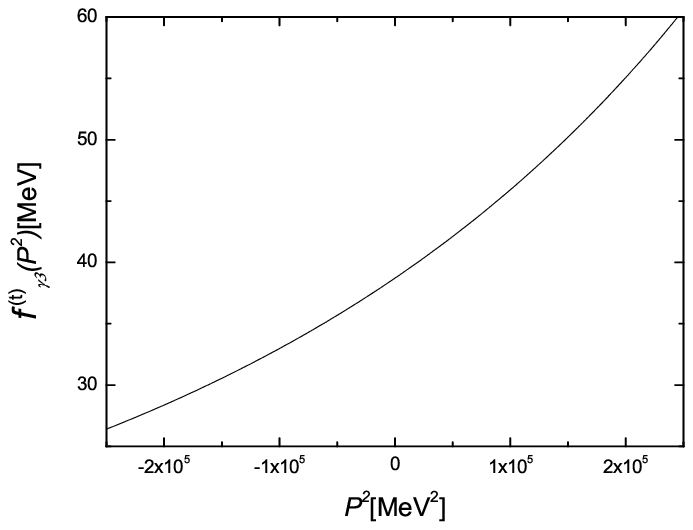}}
\end{minipage}
\hfill
\begin{minipage}{0.45\linewidth}
\centerline{\includegraphics{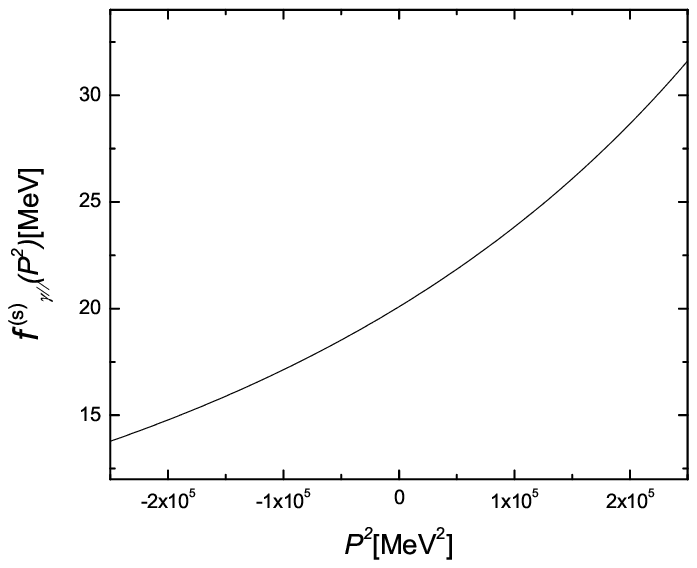}}
\end{minipage}
\caption{\label{fig:oddcoupl}The chiral-odd couplings versus $P^2$ for n=1 and $\Lambda=850\mathrm{MeV}$.}
\end{figure*}

\begin{figure*}
\begin{minipage}{0.45\linewidth}
\centerline{\includegraphics{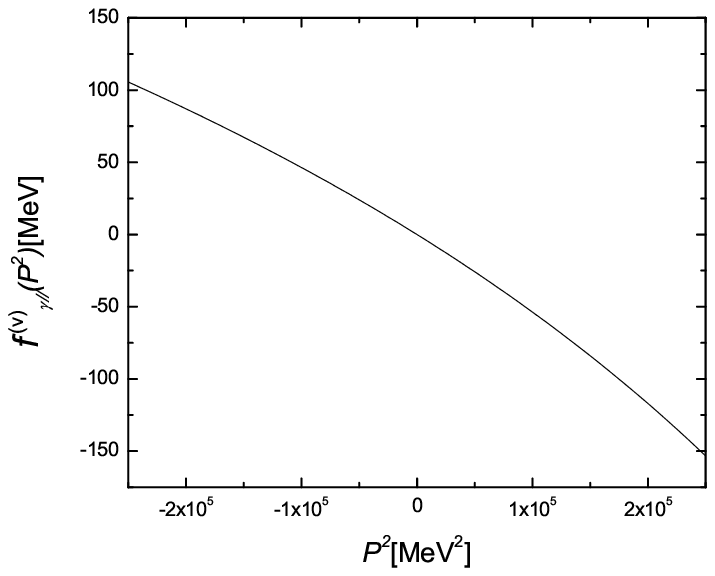}}
\end{minipage}
\begin{minipage}{0.45\linewidth}
\centerline{\includegraphics{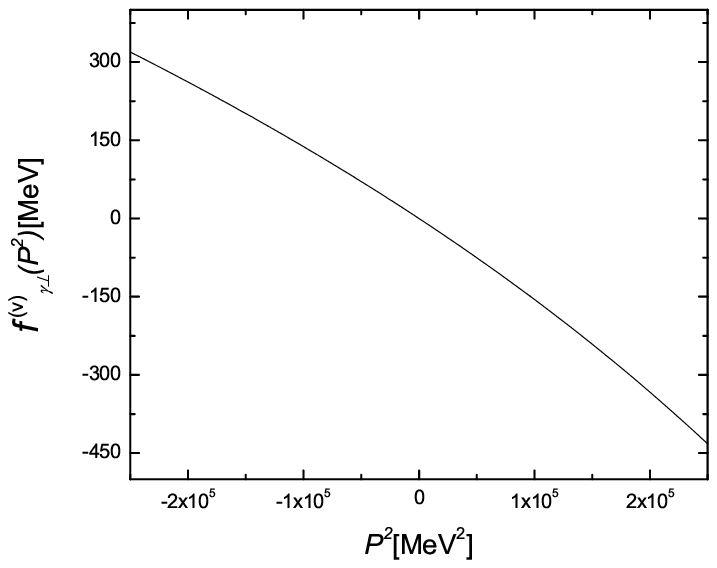}}
\end{minipage}
\begin{minipage}{0.45\linewidth}
\centerline{\includegraphics{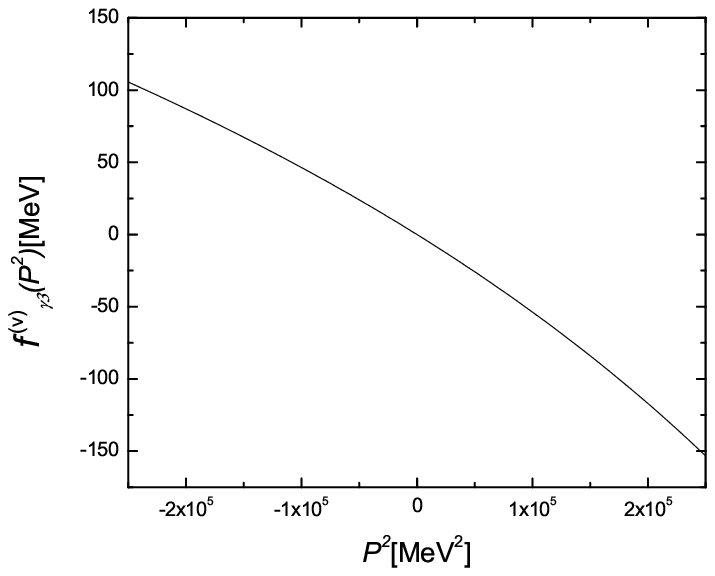}}
\end{minipage}
\hfill
\begin{minipage}{0.45\linewidth}
\centerline{\includegraphics{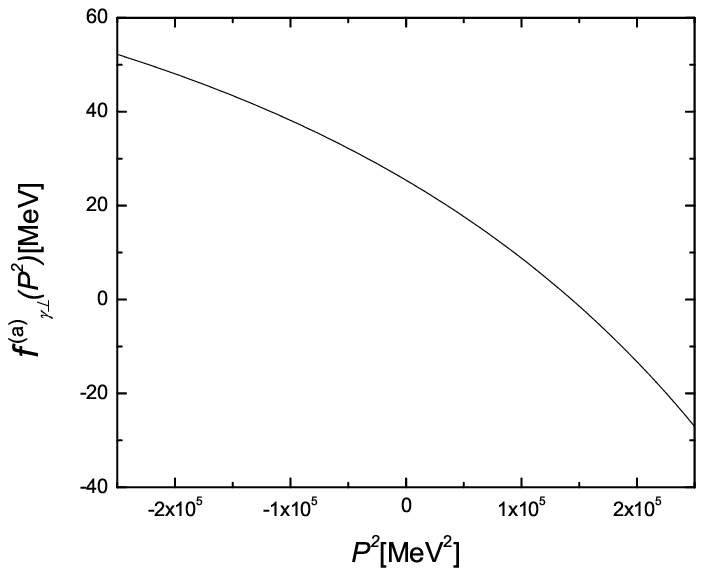}}
\end{minipage}
\caption{\label{fig:evencoupl}The chiral-even couplings versus $P^2$ for n=1 and $\Lambda=850\mathrm{MeV}$.}
\end{figure*}

From \eqref{eq:69}-\eqref{eq:72} and \eqref{eq:85}-\eqref{eq:88}, the dependencies of the couplings constants for both chiral-odd and chiral-even cases on the photon virtuality $P^2$ are displayed in Fig.\ref{fig:oddcoupl} and Fig.\ref{fig:evencoupl}. From these figures, one can see that all coupling constants are monotonic functions of $P^2$, which are increasing for chiral-odd case, and decreasing for chiral-even case. It is noticed that all coupling constants for chiral-odd case and the coupling constant $f^{(a)}_{\gamma3\perp}$ are obvious non-vanishing at $P^2=0$, while the coupling constants for the vector case are exactly zero at
$P^2=0$. All coupling constants are finite for finite $P^2$, which guarantees that only two physical transverse light-cone photon DAs,
$\phi^{(t)}_{\gamma\perp}$ and $g^{(a)}_{\gamma\perp}$, survive to be non-vanishing at $P^2=0$, while others
decouple automatically from the corresponding quark-antiquark currents.

\section{conclusion and discussion}\label{sec7}
In the present paper, we have studied systematically the off-shell light-cone photon DAs and the corresponding coupling constants for both chiral-odd and chiral-even cases up to twist-four in the instanton vacuum model of QCD. The obtained main results are:

(1) The transition matrix elements of the gauge-invariant nonlocal quark-antiquark currents sandwiched between an off-shell photon state and the vacuum are decomposed into superpositions of various Lorentz structures with the coefficients, which define the light-cone photon DAs and the corresponding coupling
constants, just based on the principle of the Lorentz covariance\cite{yuran06prd}. This formalism, in fact, parallels to the case of $\rho$ meson\cite{Ball98}. It is obvious that in this formalism only two transversal DAs,  $\phi^{(t)}_{\gamma\perp}(u,P^2)$ and $g^{(a)}_{\gamma\perp}(u,P^2)$, survive to be non-vanishing in the on-shell limit, and the others decouple automatically from the corresponding quark-antiquark currents as expected.

(2) After transferring the transition matrix elements into correlation functions, and applying the projection procedure, the various individual photon DA multiplied by its corresponding coupling constant is expressed in terms of the correlation functions. This means that we have solved the coupled equations \eqref{eq:12}-\eqref{eq:15} for the photon DAs and their coupling constants when the correlation functions are known, and these solutions are universal in the sense that their validity is independent of the specific dynamical model adopted in calculation.

(3) After choosing an appropriate gauge, for example the fixed-point gauge, where the gauge-link becomes a unit operator, we are in a position to express the leading order of the correlation functions in terms of the spectral densities of the effective quark propagator. An important point is that this quark propagator is derived from the instanton vacuum model of QCD in the singular gauge of gauge potential, which obeys the fixed-point gauge too as checked in the same way as \cite{zhang06cpl,wen11jpg}.

(4) Completing the integrations with the help of Lorentz covariance and a straightforward manipulation, we obtain the explicit analytical forms for the light-cone photon DAs and their coupling constants, and then display the dependence of the photon DAs on the momentum fraction $u$ carried by the quark for various photon virtualities $P^2$, and the dependence of the corresponding coupling constants on $P^2$. It is important to note that all the light-cone photon DAs are regular functions of $u$ and $P^2$, and their corresponding coupling constants are regular functions of $P^2$ as well. The only cutoff $v$, introduced to evaluate the integrals disappears in the final expressions, as expected from our first experience\cite{yuran06prd} where the integral for obtaining the coupling constant is already regular by choosing an appropriate integration contour. In this paper we show that this behavior is a universal feature of the integrals for all coupling constants. In this sense our treatment may be considered to be consistent.

Some points to be discussed are listed below in order:

(1) It is noticed that there are eight off-shell light-cone photon DAs appearing in our formalism. In particular, the photon DA corresponding to the scalar quark-antiquark current exists in the case $P^2\not=0$ and/or $z\not=0$, and vanishes
in the case of $P^2=0$ and/or $z=0$. The latter characteristic can be seen directly from the definition \eqref{eq:14}, and is naturally expected because there is no scalar real photon, and the local scalar quark-antiquark current is rotational invariant, and the corresponding matrix element should be zero due to the triangle rule of the angular momentum addition. However, for the former, the nonlocal scalar quark-antiquark current is not rotational invariant because the rotation operator does not commute with the complicated operator structure along the gauge link with two separated points $z$ and $-z$, and thus the corresponding matrix element dose not necessarily vanish, as shown by the explicit calculation.

(2) Like the other studies\cite{yuran06prd,prd1999pi,prd2006,prd2010}, almost all light-cone photon DAs are not vanishing at the end points except the two DAs corresponding to the scalar and axial vector case with the assumed boundary conditions. This may be a common phenomenon in a model without confinement, such as the instanton vacuum model of QCD. As pointed in \cite{npapirner}, based on a Hamiltonian containing confinement, the configurations, where one of quarks takes all of the
longitudinal momentum and the other is at rest, are expected to be suppressed.

(3) The quarks will propagate near the light-cone, $x^2\sim0$, if the off-shell photon momentum becomes minus infinity, $P^2\to-\infty$. This fact means that at the limit of $P^2\to-\infty$, the main contribution to the correlation functions comes from the asymptotic part of the propagator, and
the nonlocal quark-antiquark currents degenerate into the corresponding local ones, $z\to0$. Therefore, at that limit, \eqref{eq:12} and \eqref{eq:13} reduce to \eqref{eq:3} and \eqref{eq:4} respectively.
This asymptotic behavior leads to the consequence that the tensor and vector coupling constants tend to be equal with each other for $P^2\to-\infty$. This tendency can really be seen in Fig.\ref{fig:oddcoupl} and Fig.\ref{fig:evencoupl} respectively.

(4) For the leading twist tensor photon DA, $\phi^{(t)}_{\gamma\perp}(u,P^2)$, our result is the same as the ones derived from low-energy theory\cite{yuran06prd,prd1999pi,prd2006,prd2010} but different from the prediction asymptotic form\cite{npb2003photon} even in the real case. The reason may be that our photon DAs are
applicable to the low-energy scale, while the asymptotic ones are calculated at high-energy scale where the conformance of p-QCD is valid.

The results of our photon DAs and the corresponding coupling constants are obviously different with those in Ref. \cite{prd2010}, where the semi-bosonized Nambu Jona-Lasinio model in a nonlocal generalized form is adopted. In particular, the extra $\delta$-type singularity of both the photon DAs and the coupling constants in Ref. \cite{prd2010} does not appear in our results.

(5) As noticed already in the introduction that the dynamically generated quark mass $MF^2(k)$ in the effective quark propagator \eqref{eq:34} is momentum-dependent. This is a reflection of the non-locality of the instanton vacuum of QCD. As a consequence, the local $U_{\mathrm{em}}(1)$ gauge invariance is violated. A question is then naturally raised: to what extant our results in this paper are valid, or in other words, why the contribution due to the non-locality can be neglected? To answer it, we consider a simple minimally-local part of the effective quark propagator \eqref{eq:34}, namely by freezing the momentum squared in the form factor to be zero, and estimate the difference between the two photon DAs associated with the non-local theory and the local one as well as the difference between the corresponding coupling constants. The result is shown in Appendix C, where we have found that the derivation of both photon DA and coupling constant determined by the local theory from that in the non-local theory used in this paper is small.
This indicates that the interaction of the electromagnetic field with the quarks is indeed dominated by the local or pointlike interaction derived from the kinetic term in the effective Lagrangian. The non-local or non-pointlike coupling coming from the momentum-dependent quark mass is in fact parametrically suppressed.

We note here that an other local version of \eqref{eq:34}, for example by freezing the space-like momentum squared in $F(k)$ to be a non-zero constant $k^2\rightarrow -\mu^2$, is also allowed, and it is unclear for us to know how to define the maximally-local part of the effective quark propagator \eqref{eq:34}. Therefore, the amount of the derivation from locality shown in Appendix C may simply be considered as the upper bound.

In fact, as analyzed in the introduction, the form factor $F(k)$ \eqref{eq:33} is suppressed in $(M(k^2=-\mu^2)\bar{\rho})^2$ for $\mu^2\ll 1/\bar{\rho}$. From this reason, we stress that the local gauge invariance of the electromagnetic interaction is approximately fulfilled at least at the leading order of $M(k^2=-\mu^2)\bar{\rho}$, and we can work with the local or pointlike electromagnetic current throughout in this paper.

\section*{Acknowledgements}
This work is supported by the National Natural Science Foundation of China under Grant No. 10775105, BEPC National Laboratory Project R\&D and BES Collaboration Research Foundation, and the projects of Wuhan University of China under the Grant No. 201103013 and 9yw201115.

\appendix
\section{Photon DAs in terms of correlation functions}\label{app:aqqA}
The photon DA can be expressed in the representation of $\tau$, instead of $u$,  defined by
\begin{align}\label{eq:a1}
\phi^{(i)}_{\gamma}(\tau,P^2)=\int^1_0\ud u e^{i\xi p\cdot z}\phi^{(i)}_{\gamma}(u,P^2),
\end{align}
in terms of the correlation functions projected on the projection tensors.

Consider the tensor case, the projection tensors we need should be of order three,  and antisymmetric in the Lorentz indices $\mu$ and $\rho$ due to the same symmetric property in the corresponding correlation function. It is
obvious that there are only three different independent projection $3$-tensors being antisymmetric in $\mu$ and $\rho$ constructed from $p$ and $z$
(the other independent degree of freedom, $e^{(\lambda)}_{\perp\mu}$, disappears due to averaging the polarizations),
\begin{align*}
t_{(1)}^{\mu \nu \rho}&=\frac{1}{2}
(p^{\nu}z^{\mu}p^{\rho}-p^{\nu}z^{\rho}p^{\mu}),\\
t_{(2)}^{\mu \nu \rho}&=\frac{1}{2}(g^{\nu\mu}z^{\rho}-g^{\nu\rho}z^{\mu}),\\
t_{(3)}^{\mu \nu \rho}&=\frac{1}{2}(g^{\nu\mu}p^{\rho}-g^{\nu\rho}p^{\mu})
\end{align*}
Contracting $\Pi_{\nu\mu\rho}^{(T)}$ with $t_{(i)}^{\mu \nu \rho}$, respectively, give rise to
\begin{align}
t_{(1)}^{\mu \nu \rho}\Pi_{\nu\mu\rho}^{(T)}=&-i
\frac{P^2}{2}(p\cdot z)f^{(t)}_{\gamma\parallel}(P^2)
h^{(t)}_{\gamma\parallel}(\tau,P^2)\label{eq:a2}\\
t_{(2)}^{\mu \nu \rho}\Pi_{\nu\mu\rho}^{(T)}=&-i
(p\cdot z)\left[f^{(t)}_{\gamma\parallel}(P^2)h^{(t)}_{\gamma\parallel}(\tau,P^2)\right.\nonumber\\
&\left.+2f^{(t)}_{\gamma\perp}(P^2)\phi^{(t)}_{\gamma\perp}(\tau,P^2)\right]\label{eq:a3}\\
t_{(3)}^{\mu \nu \rho}\Pi_{\nu\mu\rho}^{(T)}=&-i
\frac{P^2}{2}\left[f^{(t)}_{\gamma\parallel}(P^2)h^{(t)}_{\gamma\parallel}(\tau,P^2)\right.\nonumber\\
&\left.+2f^{(t)}_{\gamma3}(P^2)h^{(t)}_{\gamma3}(\tau,P^2)\right]
\label{eq:a4}
\end{align}
We note here that there is, in fact, another projection $3$-tensor
\begin{align}
t_{(4)}^{\mu \nu \rho}=\frac{1}{2}
(z^{\nu}z^{\mu}p^{\rho}-z^{\nu}z^{\rho}p^{\mu})
\label{eq:a5}
\end{align}
obeying the requirement.
However, using the identity \eqref{eq:6} and the the transverse character of the correlation function  \eqref{eq:18}, the role of $t_{(4)}^{\mu \nu \rho}$ is equivalent to that of $t_{(1)}^{\mu \nu \rho}$.

For the vector case, the projection tensors should be of order two. One see that there are only three independent tensors $2$-tensors which can be
constructed from $p$ and $z$ (the other independent degree of freedom, $e^{(\lambda)}_{\perp\mu}$, disappears due to averaging the polarizations),
namely $g^{\nu\mu}$, $p^{\nu}z^{\mu}$ and $p^{\nu}p^{\mu}$.

Contracting $\Pi_{\nu\mu\rho}^{(V)}$ with these three $2$-tensors, respectively, give rise to
\begin{align}
g^{\nu\mu}\Pi_{\nu\mu}^{(V)}=&-\frac{1}{2}M \left[f^{(v)}_{\gamma\parallel}(P^2)
\phi_{\gamma\parallel}^{(v)}(\tau,P^2)\nonumber\right.\\
&\left.+4f^{(v)}_{\gamma\perp}(P^2)g^{(v)}_{\gamma\perp}(\tau,P^2)\nonumber\right.\\
&\left.+f^{(v)}_{\gamma3}(P^2)
g^{(v)}_{\gamma3}(\tau,P^2)\right]\label{eq:a6},\\
p^{\nu}z^{\mu}\Pi_{\nu\mu}^{(V)}&=-M\frac{p\cdot z}{2}f^{(v)}_{\gamma\parallel}(P^2)
\phi_{\gamma\parallel}^{(v)}(\tau,P^2)\label{eq:a7}\\
p^{\nu}p^{\mu}\Pi_{\nu\mu}^{(V)}&=\frac{1}{4}M
P^2f^{(v)}_{\gamma3}(P^2)
g_{\gamma3}^{(v)}(\tau,P^2)\label{eq:a8}.
\end{align}
We note that there are, in fact, another two projection $2$-tensors, $z^{\nu}z^{\mu}$ and $z^{\nu}p^{\mu}$, which effects are equivalent to
those of $p^{\nu}z^{\mu}$ and $p^{\nu}p^{\mu}$ respectively, by considering
\eqref{eq:6} and \eqref{eq:18} again.

Further, contracting $\Pi_{\nu}^{(S)}$ with $p^{\nu}$ (or equivalently $z^{\nu}$) leads to
\begin{align}\label{eq:a9}
p^{\nu}\Pi_{\nu}^{(S)}=-\frac{i}{2}(p\cdot z)P^2 f^{(s)}_{\gamma\parallel}(P^2)
h^{(s)}_{\gamma\parallel}(\tau,P^2)
\end{align}
for the scalar case, and finally contracting $\Pi_{\nu\mu}^{(A)}$ with the tensor $\epsilon^{\mu\nu\rho\sigma}p_{\rho}z_{\sigma}$ leads to
\begin{align}\label{eq:a10}
\epsilon^{\mu\nu\rho\sigma}p_{\rho}z_{\sigma}\Pi_{\nu\mu}^{(A)}=2M (p\cdot z)^2f^{(a)}_{\gamma\perp}(P^2)g^{(a)}_{\gamma\perp}(\tau,P^2),
\end{align}
for the axial vector case.

Up to now, we have obtained a series of the linear independent equations, namely
\eqref{eq:a2}-\eqref{eq:a4}, \eqref{eq:a6}-\eqref{eq:a8} and \eqref{eq:a9}-\eqref{eq:a10}, to determine the eight light-cone photon DAs based on the Lorentz covariance.

To solve the series of equations to get the photon DAs in terms of correlation functions, we apply the inverse Fourier transform
\begin{align}\label{eq:a11}
F[\phi^{(i)}_{\gamma}(\tau,P^2)]
=\frac{p^+}{\pi}\int d\tau e^{i2u p^+\tau}\phi^{(i)}_{\gamma}(\tau,P^2)
=\phi^{(i)}_{\gamma}(u,P^2),
\end{align}
to both sides of each one of the series of equations, and obtain
\begin{align}
f^{(t)}_{\gamma\parallel}(P^2)h^{(t)}_{\gamma\parallel}(u,P^2)
&=F\left[-i\frac{2t_{(1)}^{\mu\nu\rho}\Pi_{\nu\mu\rho}^{(T)}}{P^2(p\cdot z)}\right]\label{eq:a12}\\
f^{(t)}_{\gamma\perp}(P^2)\phi^{(t)}_{\gamma\perp}(u,P^2)
&=F\left[-i\hat{t}_{(2)}^{\mu\nu\rho}
\Pi_{\nu\mu\rho}^{(T)}\right] \label{eq:a13}\\
f^{(t)}_{\gamma 3}(P^2)h^{(t)}_{\gamma 3}(u,P^2)
&=F\left[-i\hat{t}_{(3)}^{\mu\nu\rho}\Pi_{\nu\mu\rho}^{(T)}\right]\label{eq:a14}
\end{align}
with
\begin{align*}
\hat{t}_{(2)}^{\mu\nu\rho}&=\frac{1}{2(p\cdot z)}\left[
t_{(2)}^{\mu\nu\rho}-\frac{2t_{(1)}^{\mu\nu\rho}}{P^2}\right]\\
\hat{t}_{(3)}^{\mu\nu\rho}&=\frac{1}{P^2}\left[
t_{(3)}^{\mu\nu\rho}-\frac{t_{(1)}^{\mu\nu\rho}}{p\cdot z}\right]
\end{align*}
for the tensor case, and
\begin{align}
f^{(v)}_{\gamma\perp}(P^2)g^{(v)}_{\gamma\perp}(u,P^2)
&=F[\hat{v}_{(1)}^{\nu\mu}\Pi_{\nu\mu}^{(V)}]\label{eq:a15}\\
f^{(v)}_{\gamma\parallel}(P^2)\phi_{\gamma\parallel}^{(v)}(u,P^2)
&=F\left[\frac{-2p^{\mu}n^{\nu}\Pi_{\nu\mu}^{(V)}}{M(p\cdot n)}\right]\label{eq:a16}\\
f^{(v)}_{\gamma3}(P^2)g_{\gamma3}^{(v)}(u,P^2)
&=F\left[\frac{4p^{\nu}p^{\mu}\Pi_{\nu\mu}^{(V)}}{M P^2}\right]\label{eq:a17}
\end{align}
with
\begin{align*}
\hat{v}_{(1)}^{\nu\mu}=
-\frac{1}{4M}\left(2g^{\nu\mu}
-\frac{2p^{\mu}z^{\nu}}{p\cdot z}
+\frac{4p^{\nu}p^{\mu}}{P^2}\right)
\end{align*}
for the vector case, and
\begin{align}
f^{(s)}_{\gamma\parallel}(P^2)h^{(s)}_{\gamma\parallel}(u,P^2)
&=F\left[i\frac{2p^{\nu}\Pi_{\nu}^{(S)}}{P^2(p\cdot z)}\right]\label{eq:a18}\\
f^{(a)}_{\gamma\perp}(P^2)g^{(a)}_{\gamma\perp}(u,P^2)
&=F\left[\frac{\epsilon^{\mu\nu\rho\sigma}p_{\rho}z_{\sigma}
\Pi_{\nu\mu}^{(A)}}{2M (p\cdot z)^2}\right]\label{eq:a19}
\end{align}
for the scalar and axial vector cases, respectively.
\section{Integrals $I_1$ and $I_2$ }\label{app:appB}
Consider the integral $I_1$ and $I_2$ defined in \eqref{eq:55} and \eqref{eq:56}, in which integrand involves zero and one powers of $k^{\mu}$ respectively. By
introducing the dimensionless variables
\begin{displaymath}
\eta=\frac{p^+k^-}{\Lambda^2},t=\frac{|k_{\perp}|^2}{\Lambda^2},s=\frac{P^2}{\Lambda^2},
\end{displaymath}
and completing the integration of $\delta(k^+-up^+)$, $I_1$ and $I_2$ can be expressed as
\begin{align}
I_1&=\int\frac{\ud\eta\ud t}{2(2\pi)^3}\frac{1}
{(\bar{u}s-\bar{u}\eta-t-\omega)(u\eta-t-\mu)}\label{eq:b1},\\
I_2&=\int\frac{\ud\eta\ud t}{4(2\pi)^3}\frac{\eta}
{(\bar{u}s-\bar{u}\eta-t-\omega)(u\eta-t-\mu)}\label{eq:b2}.
\end{align}
The denominator appearing in the R.H.S. of  \eqref{eq:b1} or \eqref{eq:b2} has two poles which may be assumed to be of the forms
\begin{align}
\eta_1&=s-\frac{t+\omega}{\bar{u}}\label{eq:b3},\\
\eta_2&=\frac{t+\mu}{u}\label{eq:b4}.
\end{align}
By means of the identity \eqref{eq:65}, for the partial fraction summation in the numerical simulation,
a straightforward manipulation leads to
\begin{align}
I_1&=\int\frac{\ud\eta\ud t}{2u\bar{u}(2\pi)^3}\frac{1}{(\eta-\eta_1)(\eta-\eta_2)}\nonumber\\
&=i\int\frac{\ud t}{2(2\pi)^2}\frac{1}{-u\bar{u}s+t+u\omega+\bar{u}\mu}\nonumber\\
&=\frac{i}{2(2\pi)^2}\ln(1+\frac{v}{-u\bar{u}s+u\omega+\bar{u}\mu})\label{eq:b5}
\end{align}
\begin{align}
I_2&=-\int\frac{\ud\eta\ud t}{4u\bar{u}(2\pi)^3}\frac{\eta}{(\eta-\eta_1)(\eta-\eta_2)}\nonumber\\
&=-\frac{i\pi}{4u\bar{u}}\int\frac{\ud t}{(2\pi)^3}\frac{\eta_1+\eta_2}{\eta_1-\eta_2}\nonumber\\
&=-\frac{i\pi}{2}\int\frac{\ud t}{(2\pi)^3}(u-\bar{u}+\frac{\omega-\mu-\bar{u}s}{-u\bar{u}s+t+u\omega+\bar{u}\mu})\nonumber\\
&=\frac{i}{4(2\pi)^2}(\mu-\omega+\bar{u}s)\ln(1+\frac{v}{-u\bar{u}s+u\omega+\bar{u}\mu})\nonumber\\
&+\frac{i}{4(2\pi)^2}(\bar{u}-u)v      \label{eq:b6}
\end{align}
\section{An upper bound for the derivation from locality}\label{app:appC}
\begin{figure}[!h]
\centerline{\includegraphics{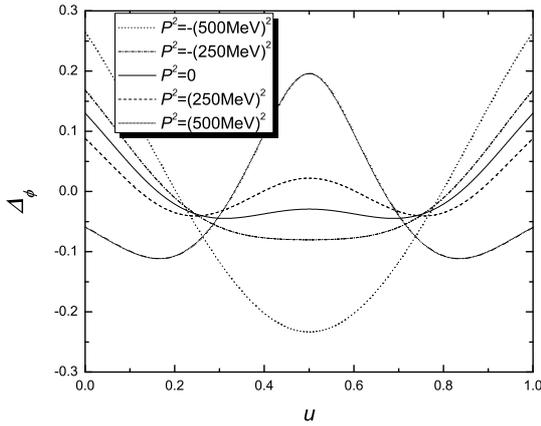}}
\caption{\label{fig:Delta-phi}The derivation of $\phi^{(t)}_{local}(u,P^2)$ from the ``local'' one.}
\end{figure}

\begin{figure}[!h]
\centerline{\includegraphics{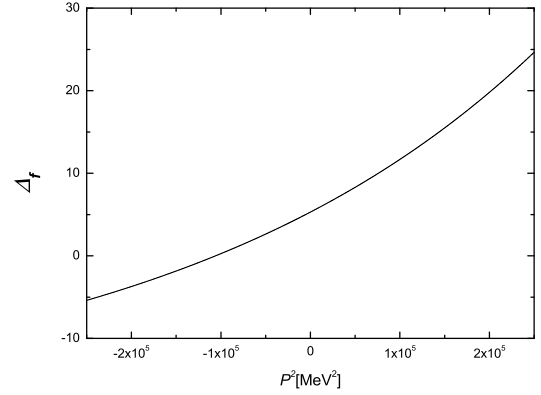}}
\caption{\label{fig:Delta-f}The derivation of coupling constant from the ``local'' one.}
\end{figure}

A simple minimally-local version of the the effective quark propagator \eqref{eq:34} is of the form
\begin{equation}\label{eq:c1}
S_F(k)=\frac{\slashed{k}+M}{k^2-M^2+i\epsilon}.
\end{equation}
Take the twist-two and chiral-odd light-cone photon DA as an example. The counterparts of $\phi^{(t)}_{\gamma\perp}$ and $f^{(t)}_{\gamma\perp}$, namely $\phi^{(t,\mathrm{local})}_{\gamma\perp}$ and $f^{(t,\mathrm{local})}_{\gamma\perp}$ associated with the local version of the propagator, can be calculated in a similar way.
Define the following three derivations of the quantities in this paper from their local ones
\begin{align}
&\Delta_{\phi}=\frac{\phi^{(t,\mathrm{local})}_{\gamma\perp}(u,P^2)-\phi^{(t)}_{\gamma\perp}(u,P^2)}
{\phi^{(t)}_{\gamma\perp}(u,P^2)}\label{eq:c2}\\
&\Delta_{f}=f^{(t,\mathrm{local})}_{\gamma\perp}(P^2)-f^{(t)}_{\gamma\perp}(P^2)\label{eq:c3}
\end{align}
The curves of $\Delta_{\phi}$ and $\Delta_{f}$ are shown in Figs.\ref{fig:Delta-phi} and \ref{fig:Delta-f}, respectively. From these two curves and the first figure in Fig.\ref{fig:oddcoupl},
it is obvious that the non-local part of the dynamical quark mass contributes within an amount of 30 percentages to the total one for both DA and the correspond coupling, and its effects are suppressed.


\bibliographystyle{unsrt}
\bibliography{photonDA}
\end{document}